\UseRawInputEncoding 
\documentclass[10pt,journal,compsoc]{IEEEtran}

\usepackage{graphicx}
\usepackage[hyphens]{url}
\usepackage{amsmath,amssymb,amsfonts}
\usepackage[font=small,tableposition=top]{caption}
\usepackage[font=footnotesize]{subcaption}
\usepackage{soul,color}
\usepackage{booktabs}
\soulregister\cite7
\soulregister\ref7
\soulregister\pageref7
\usepackage{algorithm}
\usepackage[noend]{algpseudocode}
\usepackage{tikz}
\usetikzlibrary{arrows,shadows}
\usepackage{pgf-umlsd}
\usepackage{xcolor}
\usepackage{graphics}
\usepackage{epstopdf}
\DeclareSymbolFontAlphabet{\mathrm}    {operators}
\DeclareSymbolFontAlphabet{\mathnormal}{letters}
\DeclareSymbolFontAlphabet{\mathcal}   {symbols}
\DeclareMathAlphabet      {\mathbf}{OT1}{cmr}{bx}{n}
\DeclareMathAlphabet      {\mathsf}{OT1}{cmss}{m}{n}
\DeclareMathAlphabet      {\mathit}{OT1}{cmr}{m}{it}
\DeclareMathAlphabet      {\mathtt}{OT1}{cmtt}{m}{n}

\DeclareSymbolFont{operators}   {OT1}{cmr} {m}{n}
\DeclareSymbolFont{letters}     {OML}{cmm} {m}{it}
\DeclareSymbolFont{symbols}     {OMS}{cmsy}{m}{n}    
\usepackage{hyperref}
\ifCLASSINFOpdf
\else
\fi

\begin{document}
%
%
\title{Graph-based Heuristic Solution for Placing Distributed Video Processing Applications on Moving Vehicle Clusters}
\author{\IEEEauthorblockN{Kanika Sharma, Bernard Butler, Brendan Jennings} \\      
\IEEEauthorblockA{Waterford Institute of Technology, Ireland \\
kanika\_sharma@ieee.org, bbutler@ieee.org,  bjennings@ieee.org}
}
\maketitle

\begin{abstract}
 
 Vehicular fog computing (VFC) is envisioned as an extension of cloud and mobile edge computing to utilize the rich sensing and processing resources available in vehicles. We focus on slow-moving cars that spend a significant time in urban traffic congestion as a potential pool of on-board sensors, video cameras and processing capacity. For leveraging the dynamic network and processing resources, we utilize a stochastic mobility model to select nodes with similar mobility patterns. We then design two distributed applications that are scaled in real-time and placed as multiple instances on selected vehicular fog nodes. We handle the unstable vehicular environment by a), Using real vehicle density data to build a realistic mobility model that helps in selecting  nodes for service deployment  b), Using community-detection algorithms for selecting a robust vehicular cluster using the predicted mobility behaviour of vehicles. The stability of the chosen cluster is validated using a graph centrality measure, and c), Graph-based placement heuristics are developed to find optimal placement of service graphs based on a multi-objective constrained optimization problem with the objective of efficient resource utilization. The heuristic solves an important problem of processing data generated from distributed devices by balancing the trade-off between increasing the number of service instances to have enough redundancy of processing instances to increase resilience in the service in case of node or link failure, versus reducing their number to minimise resource usage. We compare our heuristic to an integer linear program solution and a first-fit heuristic. Our approach performs better than these comparable schemes in terms of resource utilization and/or has a lesser service latency, which is a crucial requirement for safety-related applications. 
 
\end{abstract}


%
\IEEEpeerreviewmaketitle

\section{Introduction}

 The concept of Vehicular Fog computing (VFC) \cite{8641431} is derived from using the available and under-utilized vehicular resources, like in-built sensors, processors, dashboard cameras, advanced onboard devices, etc., for both crowdsensing and processing data. The VFC paradigm aims to make computation more efficient by distributed processing on the vehicular cluster, instead of offloading computation to the cloud \cite{9097892}. VFC extends the intelligence of mobile edge computing \cite{8745530}, \cite{8314696} and fog computing \cite{YOUSEFPOUR2019289} to the vehicular network, using vehicular resources to meet client requests. The vehicle nodes are treated not just as networking devices but are also used to collect a large amount of generated data and to deploy services for the fast processing of the data. This novel system of distributed service deployment reduces the traffic at the core network, saves the network bandwidth in sending local and contextual data to the Cloud, and also reduces end-to-end response latency. 

VFC also reduces the need for installing infrastructure to facilitate Intelligent Transport Systems, for improving vehicular flows and reducing congestion, for recording data for road condition monitoring to detect potholes, accidents, etc., and increasing commuter safety,  which has been theorized for more than a decade but has not been implemented \cite{8668405}. The use of Roadside Units (RSUs) or edge servers for meeting the demands of vehicular applications has also been explored, but this can result in over-provisioning of resources. These RSUs have limited coverage on the highways and have limited sojourn time with moving cars. Thus, provisioning services on RSUs can increase the cost of service migration for moving vehicles. On the other hand, the Onboard Units (OBUs) on self-driving cars have evolved in their processing capability and their ability to communicate with neighboring vehicles and keep open connections with Edge or Cloud servers. These vehicles also have a full System on a chip (SOC), for example, a Tesla car has 4 LPDDR4 RAM chips, with complete redundancy to have failure resistance for any system on board \cite{WinNT}.

For this system to operate properly, we leverage the mobility pattern of slow-moving vehicles to estimate the ongoing availability of these vehicles to perform tasks. The novel distributed services we propose are scaled in real-time and deployed on the vehicle cluster such that we get a robust initial placement with less need to reconfigure services. Our model can support many applications using crowdsourced data from vehicles. Applications include sensing applications like pedestrian detection to understand human engagement with coffee shops, gas stations, and other locations. The data is also used for safety-related applications, to detect pedestrians and cyclists to alert drivers and increase the safety of commuters. The collected data can also be used for distributed vehicle flow measurement for traffic management, as part of building Smart cities \cite{9002050}. The collected data can also be used to disseminate real-time 3D road maps for autonomous driving which requires precision data and complex data processing \cite{8946549}. As can be noted, all these applications have varying processing and latency requirements, thus our model focuses on two specific applications with different levels of computation-intensive tasks.

There are many challenges in utilizing these moving vehicle nodes and bringing the data processing services close to the source of data generation. We introduce placing distributed and flexible services instead of static services, where each computational task is broken down into smaller sub-tasks to utilize the distributed resources available in different fog nodes. Each sub-task is then scaled to multiple instances to increase resilience in the service. The estimation of available infrastructure, the scaling of microservices to the optimal number of instances, and finding the right candidate vehicle nodes for task instance placement makes the problem hard to solve.

In this paper, to effectively use the unstable and dynamic vehicular network for service provisioning, we leverage the macroscopic mobility model for intersections that capture the predictability of vehicular flows in the vehicular network, introduced in our previous work \cite{sharma2021scaling}. We used real vehicle density data to analyze peak traffic congestion patterns and then calibrate the microscopic mobility model according to real traffic density traces. We then introduce a flexible service model that can be scaled to multiple service instances in real-time, based on the application demand. We model the placement of these distributed services on moving vehicle nodes as a constrained optimization problem. We profile two applications, a data collection service and an object detection service for pedestrian detection, to understand the resource usage of these distributed applications. We use two community-detection algorithms to select nodes that have a higher probability to stay within the cluster for a longer time, reducing the need for service reconfiguration. We then introduce a graph-theory-based heuristic that promotes placing task instances optimally within the vehicle clusters. We compare our placement technique to an integer linear program solution, and to a first-fit approach based solely on reducing service latency. 

The paper is organised as follows: \S2 highlights the related research undertaken on task offloading in VFC and vehicular crowdsensing. \S3 introduces the terminology used in the paper and gives a detailed system model. In \S4 we specify the two distributed application types considered in this work. We then give details on the network topology and distributed service model along with the notations used for the mathematical modeling. \S5 covers the service scaling and placement constraints and the infrastructure constraints. The section also details the mobility model and the objective function of the constrained optimization problem. \S6 describes the community-detection-based node selection and the graph-based service placement heuristics. \S7 has a detailed evaluation of the introduced technique compared to the ILP and first-fit solution. We also show the performance of the solution in regards to the service time and the state of the selected nodes' cluster over time. The paper is concluded in \S8 where the future work has also been suggested. 


\section{Related works}
We discuss the existing task offloading and service placement schemes in VFC models. We highlight the challenges in implementing task offloading in a dynamic vehicular environment and discuss existing schemes addressing these challenges.

\subsection{Task offloading in Vehicular Fog Computing}

Task offloading schemes in VFC are widely researched and are designed to minimize processing latency \cite{8931659}, without compromising the quality of the service \cite{8489874} also focusing on efficient resource allocation \cite{8529256}. Most of these works focus on utilizing the available mobile edge computing infrastructure for carrying out compute-intensive tasks. Liu et al. \cite{9495832} have introduced a three-layer service architecture for offloading vehicular applications in vehicular fog, fog server, and the central cloud. To solve the Probabilistic Task Offloading problem they introduce an alternating direction method of multipliers (ADMMs) and particle swarm optimization (PSO), to divide the problem into multiple unconstrained sub-problems that iteratively reach an optimal solution. 

Liang et al. \cite{9415624} suggest the use of public transport facilities like buses and taxis as fog nodes to reduce the randomness of vehicle movement with fixed bus trajectories. To solve the interruption problem caused by vehicle mobility as well as the problem of delay and reliability loss, they introduced a low-latency information distribution scheme for VFC. They study network topology dynamics to evaluate and predict connection status between fog nodes and the adjacent vehicles. Qiao et al. \cite{8436044} takes advanced driver assistant systems and autonomous driving as the use case for a distributed and collaborative task offloading scheme with a guarantee of low communication and computation latency. They work on removing redundant computation tasks based on task similarity and computation capacity. Vehicles are partitioned into the task computing sub-cloudlet to provide underutilized communication and computation resources. Vehicles with lesser similarities are partitioned into the task offloading sub-cloudlet to assign their computation tasks to edge infrastructures. Sharma et al. \cite{8581027} introduced the problem of service mapping and service placement in vehicular networks as an integer linear problem with an objective of efficient network bandwidth utilization.  

Lee et al. \cite{9097892} introduced a reinforcement learning-based resource allocation model for the continuous and high dimensional action-space in a VFC environment. They use a simple vehicular mobility model for parked vehicles and a realistic mobility model based on Zurich traffic traces \cite{7056538} to determine vehicles arriving and departing from parking lots. Iqbal et al. \cite{9046146} developed a blockchain-based, distributed reputation ledger to identify malicious vehicles. They then propose a framework to handle peak workloads using nearby fog vehicles and the RSUs. The reputation score is awarded to the vehicles upon task completion to enable a decision model for task assignment.      

\subsection{Crowdsensing in vehicular networks}

Vehicular crowdsensing (VCS) utilizes the available sensing capability of vehicles and their mobility pattern to accomplish sensing tasks. The aim of a VCS system is different from a VFC system but has common challenges like modeling vehicular mobility and the need for an incentive mechanism for the VCS system. Zhao et al. \cite{9173810} derived a long-term strategy to build a deep reinforcement learning-based incentive mechanism. They model the vehicle dynamics via a dynamic radio channel with a selection of sine, piece-wise linear, and Markov-chain channel models. Edge devices are also used for detecting parking space availability. Grassi et al. \cite{10.1145/3132211.3134452} depict the feasibility of deploying image-based, machine learning techniques at the network edge. They use smart cameras placed on dashboards to capture information related to parking availability without any human intervention. Zhu et al. \cite{8879201} focuses on the challenging task of finding parking availability for autonomous vehicles. They collect parking information from crowdsensing and use VFC to estimate parking availability and inform client vehicles. 

\subsection{Role of VFC and VCS}

As pointed out in this section, there is a lot of active research in the field of VFC and VCS for many different application profiles and with different performance metrics as an objective. Our work focuses on utilizing vehicular mobility patterns in urban traffic to deploy specialized services on these vehicles. Instead of focusing on the vertical edge computing, fog computing, and cloud computing model, we focus on horizontal scaling of service and graph-based representation of vehicle clusters. We also introduce a community-based node selection and service placement scheme suitable for the scenario. 

\section{System Model}
\label{sec:System Model}

\begin{figure}[t!]
\centering
\includegraphics[width=5cm]{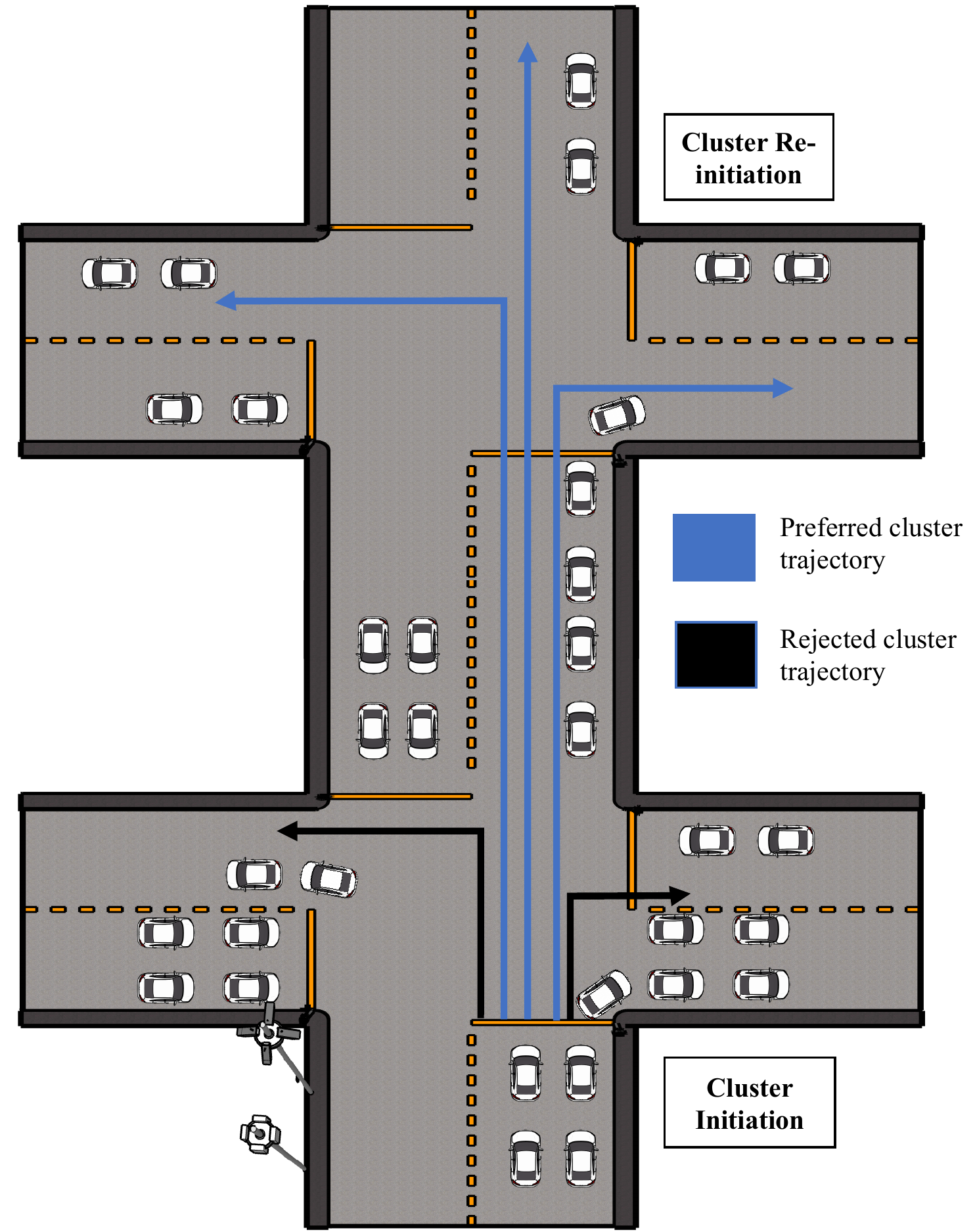}
\caption{Vehicle Clusters form, but membership changes over time. Clusters accept service chain placement requests from RSUs and perform service chain scaling and placement.}
\label{fig:scenario}
\end{figure}

\begin{figure}[t!]
\centering
\includegraphics[width=8cm,height=7cm]{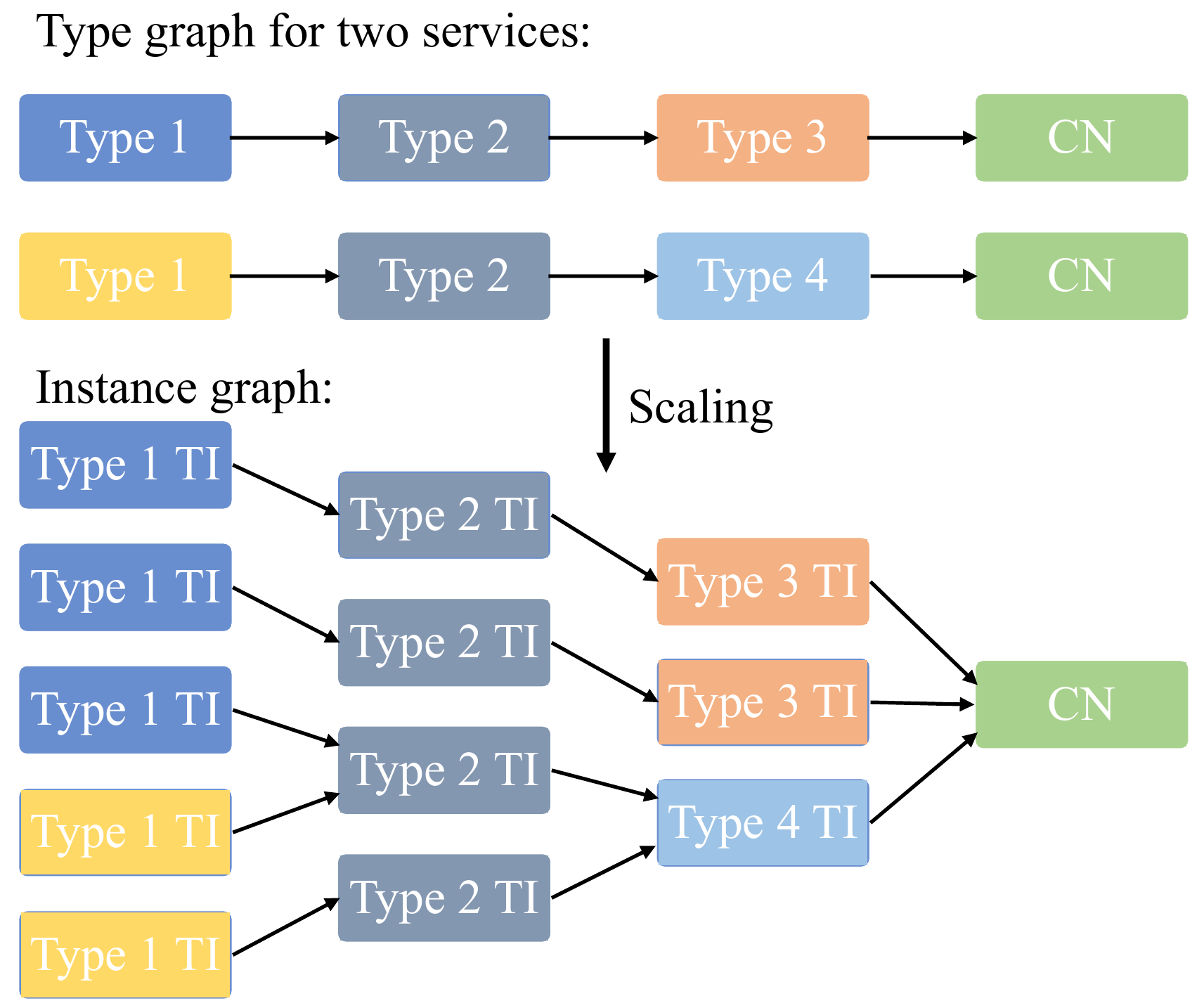}
\caption{Service model depicting tasks and their inter-dependencies. The Type graph is scaled to the Instance graph based on the resource state of the vehicle cluster.}
\label{fig:ServiceModel}
\end{figure}

This section details the proposed model for selecting a well-connected cluster and managing the placement of the services on the vehicle cluster. This placement is managed with the coordination of the cluster's Control Node (CN) and the RSUs. The RSU receives requests from clients to deploy services. The client could be a one-off vehicle node moving along with the cluster or a surveillance request from traffic authorities. In this paper, we assume that the service request is received and decomposed by the RSU in the form of a linear chain of tasks. As depicted in Fig.\ref{fig:scenario}, the RSU detects the presence of vehicles that have previously subscribed to a brokerage service and hence are prepared to lease their resources for service provisioning. The RSU also stores and updates the database of the mobility patterns of these vehicles. Each vehicle has a probability of taking a certain trajectory based on its historic mobility pattern. Based on the preferred trajectories of the cluster, each vehicle node has a certain probability of following the cluster trajectory. A weighted graph is created where each link is weighted by the probability of two nodes staying together for a duration of time, called the Cluster Cohesion Probability (CCP). Based on this graph, the most well-connected cluster is selected, and the node with the highest connectivity is elected as the CN, based on a centrality measure.

\begin{figure}[t!]\centering
\includegraphics[width=0.9\linewidth,height=9cm]{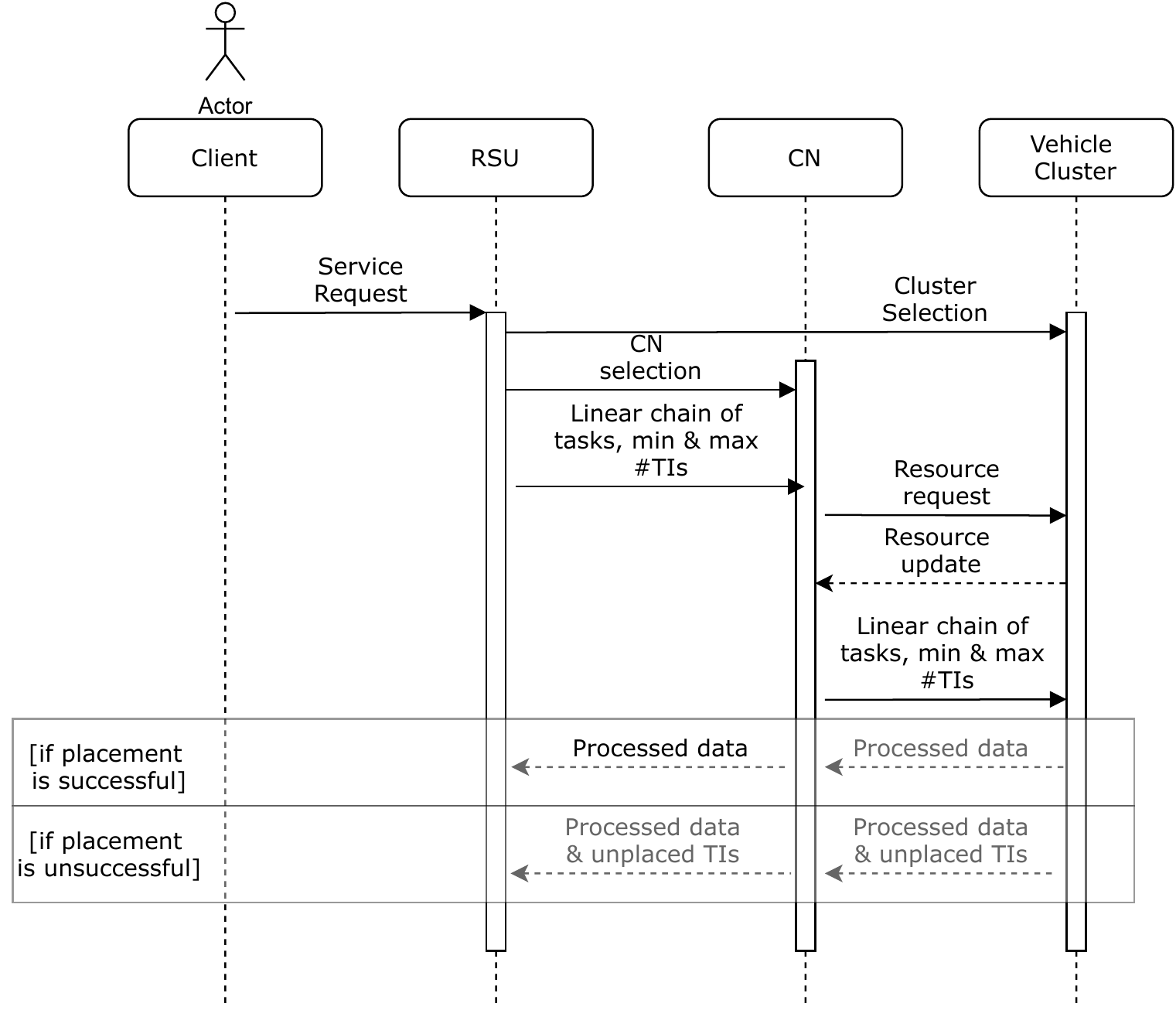}
\caption{Sequence diagram depicting the order of interactions between the client, the RSU, the CN and the vehicle cluster}
\label{fig:SequenceDiagram}
\end{figure}

Once the vehicle cluster and the CN are selected the process of service placement begins. The process of service placement and the order of interactions between the client, the RSU, the selected CN, and the vehicle cluster is represented in Fig.\ref{fig:SequenceDiagram}. The RSU sends the linear Type graph to the CN in the form of docker images. The CN also collects the updated resource information from the vehicular cluster. Initially, the minimum number of instances of each task is placed on the cluster, to process multiple data flows in parallel. Based on the number of video task instances (TIs) and the amount of processing required the heuristic scales to more processing TIs by requesting TI images from the CN. If there are no more potential nodes left to place new TIs, the collected data is sent to the nearest RSU with the remaining TIs in the linear chain left to be placed for processing. The RSU can request a re-initiated cluster to place the remaining processing TIs on it.



\subsection{Terminology}

\begin{itemize}
    \item \textbf{Vehicle Clusters}: As vehicles slow down at intersections or busy urban routes, we recruit a group of vehicles that are likely to stay together, based on their previous mobility pattern. Each vehicle subscribes to the service of leasing its resources and has a cluster cohesion probability (CCP) based on its microscopic mobility pattern. The vehicle clusters are represented as undirected graphs. 
    \item \textbf{Control Node (CN)}: The CN coordinates the vehicle cluster(s) and routes messages to/from clients/Roadside Units (RSUs) and is selected based on its betweenness centrality, which is a metric popularly used in social network analysis. For any node, the betweenness centrality is calculated as the fraction of the shortest paths between any pair of nodes in the cluster that pass through that node. From a network point of view, the node with the highest betweenness centrality is a critical point of information flow in a cluster. 
    \item \textbf{Roadside Units (RSUs)}:  The RSUs are edge devices with much higher resource capacity relative to the vehicle nodes. The RSU receives requests from clients and initiates a vehicle cluster at an intersection. The CN sends the metadata of unplaced tasks to the RSUs. The historic mobility patterns of the registered vehicles are also stored at the RSU.
    \item \textbf{Task}: Tasks are the smallest unit of a video processing service. Based on the application, tasks could have different functionality like data filtering or data compression. Tasks could also be more complex and processing-intensive like local object detection. The application types and the example tasks are described in the application section. Each task is scaled to multiple task instances (TIs) to handle multiple sources of data and to process data streams in parallel. The multiple TIs also increase the resilience of the service against node failures (when vehicles leave) and link failures (when connectivity is lost). 
    \item \textbf{Service}: Each service constitutes different types of tasks with varying functionality. The linear chain of tasks is called the Type graph. Each task can have multiple TIs as depicted in Fig.{\ref{fig:ServiceModel}}. We place two different services on the vehicle cluster, with some shared tasks (Type 2) to promote the reuse of TIs. This 'Type graph' is scaled to an 'Instance Graph' by both the ILP solver and the service scaling and placement heuristic proposed in this paper. 
\end{itemize}

\section{Model}

\subsubsection{Application Type}

We consider a linear chain of video collection and processing tasks. We place multiple video collection TIs to increase the scale of video collection which results in better accuracy for object detection applications. To process multiple video streams, we scale all the processing tasks to multiple TIs, to utilize the limited processing capacity of vehicles. The multiple TIs also increases the resilience of the service in a dynamic vehicular network where mobility of vehicles increases node and link failures. We present two distributed services that follow our distributed service model:

\begin{figure}[t!]\centering
\includegraphics[width=\linewidth,height=5cm]{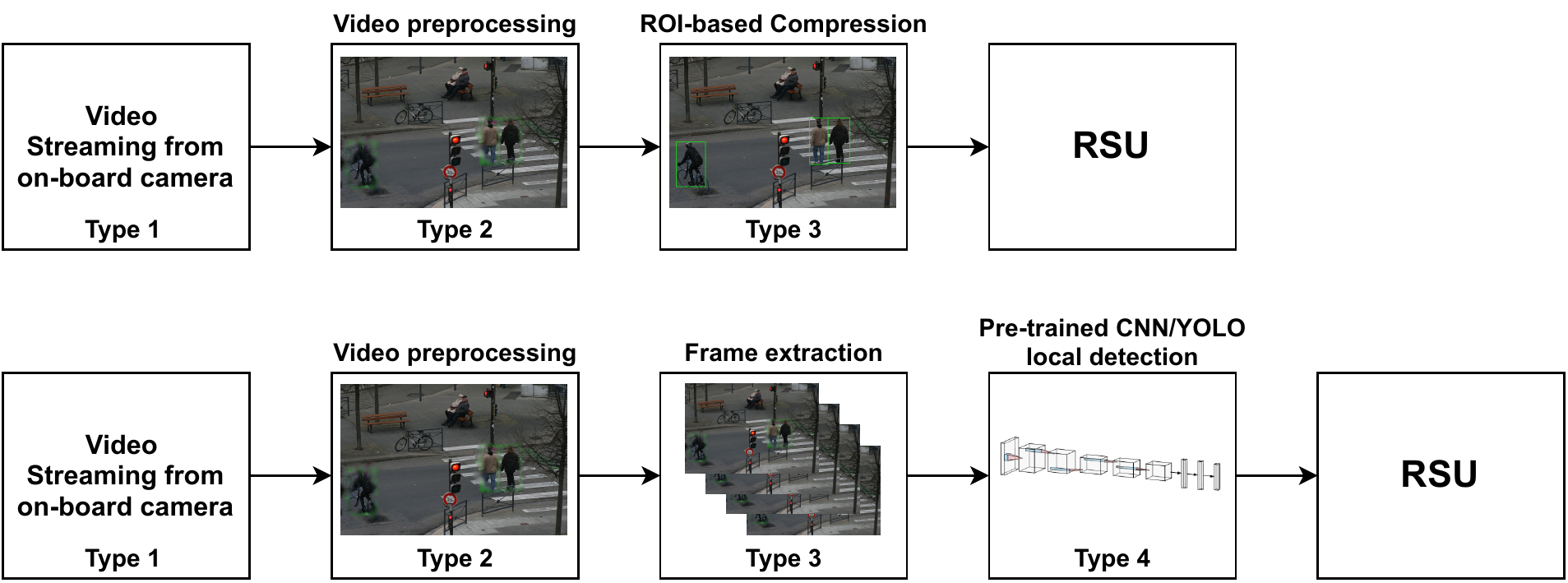}
\caption{Two types of applications of chain length 3 and 4}
\label{fig:ApplicationTypes}
\end{figure}


\begin{enumerate}
    \item \textbf{Data Collection service}: For this kind of service, an initiated cluster acts as "moving sensors" and only pre-processing and compression is carried out at the cluster. Most of the application-specific processing is carried out on more powerful edge computing nodes at RSUs. As an example, we take an application where Type 1 is a video capturing TI, Type 2 is a video pre-processing  TI. Type 3 is a video compression TI. 
    
    To profile this application, at Type 1 TI, we first capture the video using OpenCV. At Type 2 TI, the video is pre-processed using Gaussian blur and simple thresholding. For Type 3 TI, the images are compressed using RoI-based image compression. The task function, techniques and commonly-used algorithms are tabulated in Table \ref{tab:app1} for this data collection application. 
    
    
    This collected data is sent to the mobile edge for an application that requires specific data over time like traffic monitoring and traffic management. Similarly, a large amount of data can also be collected for applications that require inference from more sophisticated Deep Neural Network (DNN) models that demand powerful cloud computing devices. For example, multiple 3D road maps can be generated from multiple sources of video data. This kind of application requires real-time and local information. Our service model sends pre-processed and compressed data, reducing the overhead of data transmission, for the data-intensive map generation task that can then be executed on mobile-edge computing devices.  

\begin{table}[]
\resizebox{\columnwidth}{!}{%
\begin{tabular}{|c|c|c|c|}
\hline
\textbf{Task Type} & \textbf{Task Function}    & \textbf{Task Techniques} & \textbf{Algorithms} \\ \hline
Type 1    & Real-time video capturing &                          &                     \\ \hline
Type 2 &
  Video preprocessing &
  \begin{tabular}[c]{@{}c@{}}Image Enhancement,\\ Noise reduction\end{tabular} &
  \begin{tabular}[c]{@{}c@{}}Gaussian blur, \\ Simple thresholding\end{tabular} \\ \hline
Type 3 &
  Video compression &
  \begin{tabular}[c]{@{}c@{}}non RoI-based compression\\ RoI-based compression,\end{tabular} &
  Huffman encoding \\ \hline
\end{tabular}%
}
\caption{Task function, techniques and commonly-used algorithms for Application 1 tasks}
\label{tab:app1}
\end{table}

\begin{table}[]
\resizebox{\columnwidth}{!}{%
\begin{tabular}{|l|l|l|l|}
\hline
\textbf{Task type} & \textbf{Task function} & \textbf{Task techniques} & \textbf{Algorithms} \\ \hline
Type 1 & Video collection &  &  \\ \hline
Type 2 & Video pre-processing & \begin{tabular}[c]{@{}l@{}}Image enhancement/\\ Noise reduction\end{tabular} & \begin{tabular}[c]{@{}l@{}}PCA tranformation\\ Noise removal using\\ Wiener filter\end{tabular} \\ \hline
Type 3 & Feature Extraction & \begin{tabular}[c]{@{}l@{}}Regularization/\\ Dimensionality \\ reduction\end{tabular} & \begin{tabular}[c]{@{}l@{}}Independent \\ Component \\ Analysis\end{tabular} \\ \hline
Type 4 & Object detection & \begin{tabular}[c]{@{}l@{}}Local, pre-trained \\ model for object\\ detection\end{tabular} & Faster R-CNN, YOLO, tinyYOLO \\ \hline
\end{tabular}
}
\caption{Task function, techniques and commonly-used algorithms for Application 2 tasks}
\label{tab:app2}
\end{table}

    \item \textbf{Object Detection application}: The aim of building this distributed object detection application is to deliver a prompt and communication-efficient data processing service leveraging the resources available in moving vehicles. For this kind of application, all the processing of the collected data is executed on the vehicle cluster. Such applications have local context and scope, for example, using object-detection techniques to identify vulnerable pedestrians and alerting drivers in the vicinity. The Type 1 TIs for this application are of video collection type. The Type 2 TIs are the same pre-processing TI as Type 2 in application I. Type 3 TI is a frame extraction type that transforms video stream to images based on an extraction rate. For slow-moving pedestrians, the extraction rate is low which reduces the computation intensity of the task. For Type 4 TI, we use a pre-trained object detector called YOLO \cite{yolov3} which determines if the object of interest is in the frame, from a detectable object pool. The task function, techniques and commonly-used algorithms are summarised in Table  \ref{tab:app2} for this application type.
   
If the Type 4 TI finds unknown objects, the images can be transmitted to back-end servers, which can thereby update the inference model of local detectors, but the global knowledge and cloud involvement are not in the scope of the applications we are defining for local detection.

\end{enumerate}

We use the linear model described in \cite{8489874} to determine the memory usage for video streaming. For the case of video streaming the memory usage ranges between 110-220 MB. The data size for each frame is given in the range of 2.7-33.7 KB for five different video resolutions (1920 * 1080, 1280 * 960, 960 * 720, 640 * 480, 320 * 240). For the object detection application, we ran pre-trained YOLOv3 \cite{yolov3} model on an Linux OS system with 8GB RAM and i7-6500U running at 2.50 GHz and got a processing latency of 7.417 seconds. The detection has 16-18\% of CPU usage. We also ran Tiny YOLOv3 and got a processing latency of 0.31277 seconds with lower confidence scores. The detection uses 4-5\% of CPU and can be used on resource constrained on-board units for non-safety related applications that can afford lower accuracy.  







\subsection{Network Topology}

The network topology consists of moving nodes that halt at an intersection and the roadside unit (RSU) that receives application requests from clients. The cluster is initiated by selecting nodes based on their mobility pattern and resource availability. This information of vehicles willing to lease their resources is stored and updated at the RSU. A vehicle cluster is initiated by detecting a density of $\mathbb{I}$ nodes subscribed to provide their onboard computing and camera resources. The RSU represented as $\mathbb{I_{RSU}}$ collects mobility and resource information from the $\mathbb{I}$ subscribed nodes. The mobility of each node is presented as the cluster cohesion probability (CCP), represented as $p_{I}$, which is the probability of the node staying at a particular road segment in a time interval from $[t_1,t_2]$. We also derive the communication link probability between two vehicles as the joint CCP for two vehicles to communicate over a period of time $[t_1,t_2]$.

The selected cluster of vehicles is represented as a directed graph $G(V,E)$. Each node of the graph $i \in V$ has K resource types. The available processing capacity, for each resource type k, is represented as $C_k(i)$. Each node $i$ has a probability of staying on a road segment for a time period $[t_1,t_2]$, represented as $P_{(t_1,t_2)}$. The CCP of the RSU is equal to 1 as it is stationary and is always available from the viewpoint of mobility. The available link capacity between two nodes $i_1$ and $i_2$ is represented as $B(i_1,i_2)$ Kb/s. The joint probability between two nodes $i_1$ and $i_2$ depicts the probability of both nodes to stay together in a road segment for a period of time $[t_1,t_2]$ and is represented as $P_{t_1,t_2}(i_1,i_2)$. This is crucial for placing TIs that depend on other TIs for input data for task completion. 

\subsection{Distributed Service Model}

The service model is composed of tasks, denoted as $s_p$, each with a different processing function or type, represented as $p$. Due to the limited resource capacity in each vehicle node, a service is composed in a distributed manner as a linear chain of tasks. Due to the dynamic nature of the vehicular network, each task can be scaled to multiple TIs, represented as $s_{pj}$ where $p$ represents the type of each TI and $j$ represents the number of TIs. The number of TIs for each task $s_p$ is $N_{s_p}$ and the maximum number of allowed TIs for each type is given as $N_{s_p^{\textrm{max}}}$. The objective of scaling tasks to instances is to increase the robustness in the service model, especially because of the link and node failure due to the wireless connectivity and vehicle mobility. The resource requirement of type $k$ for each task type $p$ is represented as $D_{kp}$, where $k \in \{1,2,3,4\}$ for CPU, memory, GPU and video camera. The incoming flow from task types $s_{p_1}$ to $s_{p_2}$ is given as $F(s_{p_1},s_{p_2})$. 

The objective of this optimization is to find nodes that have a higher probability of staying together over a period of time. We then place two services of varying chain lengths (3 and 4), as described above. We model the mobile infrastructural resource constraints and the constraints required for placing a flexible and scalable service on the infrastructure. We then optimize resource usage in the service placement on the vehicular cluster through node and link cost, which is the sum of processing resources on vehicles and the communication cost for the data flow between the distributed tasks. The resource utilization is normalized to total available resources and weighted by the CCP to take into account the mobility of vehicle clusters.

\section{\textbf{Service scaling and placement constraints}}

We define the service scaling and placement problem as a constrained optimization problem. We first define the constraints for the distributed service scaling, which are given as: 

\subsubsection{\textbf{Flow capacity constraint}} The processing requirement for a flow from TI $s_{p_1j}$ to $s_{p_2j}$ is given as $C(F(s_{p_1j},s_{p_2j}))$. The constraint \ref{eq:FlowRateConstraints} ensures that each TI has enough processing capacity for the incoming flow. The constraint is given as:  
\begin{equation}
\small{
\begin{aligned}
\forall{i_1 \in \{1,\ldots,\mathbb{I}\}; i_2 \in \{1,\ldots,\mathbb{I}\}; i_1 \neq i_2}\\
\sum_{\forall{p_1,j_1; p_2,j_2; p_1 \neq p_2}} M(p_2,j,i_2)C(F(s_{p_{1}j},s_{p_{2}j})) \leq C(s_{p_{2}j})
\end{aligned}
}
\label{eq:FlowRateConstraints}
\end{equation}
where $C(s_{p_{2}j})$ is the available processing capacity at TI $s_{p_{2}j}$. Here, $M(p_2,j,i_2)$ is a binary mapping variable which is 1 when the TI $s_{p_{2}j}$ is mapped to node $i_2$ and is 0 otherwise. 

\subsubsection{\textbf{In-network processing constraint}} Constraint ~\ref{eq:FlowConservationConstraints} ensures that the flow is processed at each TI before being sent to the control node. To ensure that, we calculate the ratio of incoming to outgoing flow which should be equivalent to the data processing factor of each TI. The processing factor is given for each task type p and is given as $\alpha_{p}$. The constraint is presented as: 
\begin{equation}
\small{
\begin{aligned}
\forall{i_1 \in \{1,\ldots,\mathbb{I}\}; i_2 \in \{1,\ldots,\mathbb{I}\}; i_1 \neq i_2}\\
\sum_{\forall{p_1,j_1; p_2,j_2; p_1 \neq p_2}} F(s_{pj},s_{(p+1)j})\alpha_p \leq F(s_{(p+1)j},s_{(p+2)j})\\
\end{aligned}
}
\label{eq:FlowConservationConstraints}
\end{equation}
where $0 \leq \alpha_{p} \leq 1$ . The data processing factor is 1 for forwarding nodes as it does not process the incoming data flow. The incoming flow from $s_{pj}$ to $s_{(p+1)j}$ is given as $F(s_{pj},s_{(p+1)j})$. The outgoing flow from $s_{(p+1)j}$ to $s_{(p+2)j}$ represents the flow that has to be processed at the TI $s_{(p+2)j}$.

\subsubsection{\textbf{Service Scaling constraint}} The constraint \ref{eq:servicescalingConstraint} ensures that the TIs are scaled to the maximum number of TI specified for each task type p. This constraint also ensures that there is at least one TI for for each task type. This constraint is given as:
\begin{equation}
\small{
\begin{aligned}
\forall{p} \ \
N_{sp_{min}} \leq   N_{sp} \leq N_{sp_{max}}
\end{aligned}
}
\label{eq:servicescalingConstraint}
\end{equation}
where $N_{sp}$ is the number of TI of task type p. The maximum allowed TIs for the task type p is given as $N_{sp_{max}}$ and the minimum number of TIs for task type p is given as $N_{sp_{min}}$ which is set to 1 for our model.

\subsection{\textbf{Infrastructure constraints}}

The infrastructure constraints ensure the the node and link placement meets the resource constraints for the service placement. The infrastructure constraints are given as: 

\subsubsection{\textbf{Node Resource constraint}}

The resource requirement for a TI is represented as $D_{pk}$ where p is the type of task and k is the resource type where $k=1$ is CPU cycles requirement, $k=2$ is memory capacity requirement, k = 3 is video camera resource requirement and k = 4 is the GPU availability. A decision variable $M(p,j,i)$ is used if TI $s_{pj}$ is mapped to node $i$. The node resource capacity ensures that there is enough available capacity at a node to support a TI $s_{pj}$. The constraint is given as:

\begin{equation}
\small{
\begin{aligned}
\forall{i \in \{1,\ldots,\mathbb{I}\}, k \in \{1,\ldots,\mathbb{K}\}},\sum_{\forall p,j} M(p,j,i).D_{pj,k} \leq C_{k}(i)
\end{aligned}
}
\label{eq:NodeResourceConstraint}
\end{equation}
where $C_{k}(i)$ is the available capacity at node $i$ for resource $k$. 

\subsubsection{\textbf{Bandwidth constraint}}

The bandwidth constraint ensures that the bandwidth requirement between two TIs, given as $F(s_{pj},s_{(p+1)j})$, is less than the available bandwidth capacity over the entire path between two task instances $s_{pj}$ and $s_{(p+1)j}$. The path between two nodes $i_1$ and $i_n$ is a list of bandwidth of variable length. It stores available capacity over all forwarding nodes between $i_1$ and $i_n$, if there is no direct link between the two nodes and the available bandwidth link between the two nodes if they are directly connected. The bandwidth capacity of the path is represented as $path[B(i_{1},i_2),\dots,B(i_{n-1},i_n)]$. We enable this to support multihop clusters for cases where nodes might not be connected through a direct path but are connected over multiple hops. The constraint is given as:
\begin{equation}
\footnotesize{
\begin{aligned}
\forall{i_1 \in \{1,\ldots,\mathbb{I}\}; i_2 \in \{1,\ldots,\mathbb{I}\}; i_1 \neq i_2}\\
\sum_{\forall{p_1,j_1; p_2,j_2; p_1 \neq p_2}}  M(p_1,j_1,i_1)F(s_{pj},s_{(p+1)j})M(p+1,j_2,i_n) \\
\leq  \min(path[B(i_1,i_2),\dots,B(i_{n-1},i_n)])
\end{aligned}
}
\label{eq:BandwidthConstraints}
\end{equation}
\subsection{Mobility modeling}

Each vehicle node has a certain probability of choosing a road segment based on its historic mobility pattern. The mobility history for each vehicle node is stored in the RSU along with the time stamp. We aim to select vehicles with the highest probability of staying at the selected road segment which is $RS_j$ in our case, depicted in Fig. \ref{fig:MobilityModelling}. A transition probability matrix stores the mobility probability for different road segments for all the vehicles registered to lease their resources and participate in the crowdsourcing service. New vehicles registering for the first time are also added to the table. 

For discovering participating vehicles for the service deployment, the RSU broadcasts probe messages for participation requests. If already registered vehicles with known transition probability responds, they are given a priority over newer vehicles that want to participate. The newest participants are given the least confidence score. The confidence score is not issued according to the performance of the deployed task. It is a simple measure of updating confidence score if the vehicles follows its historic mobility trajectory. The confidence score is updated as the number of times the vehicle followed a preferred trajectory, over the total number of trips registered by the vehicle. Once the RSU updates its participants list, the RSU then runs the community detection-based node selection algorithm on a group of participants with a confidence score above a pre-decided threshold value. 

We consider each road segment to be a Markov state. The vehicle transitions in the Markov process when moving from one road segment to the next. The vehicles follow a Markov memory-less property, wherein the node transitions from state n to n+1 and is independent of state n-1. We record the transition of a vehicle from state $RS_i$ to $RS_j$ as the number of times a vehicle transition to segment $RS_j$ given the vehicle was at $RS_i$ in its previous state. The probability is given as: 

\begin{figure}[htbp]\centering
\includegraphics[width=\linewidth,height=3cm]{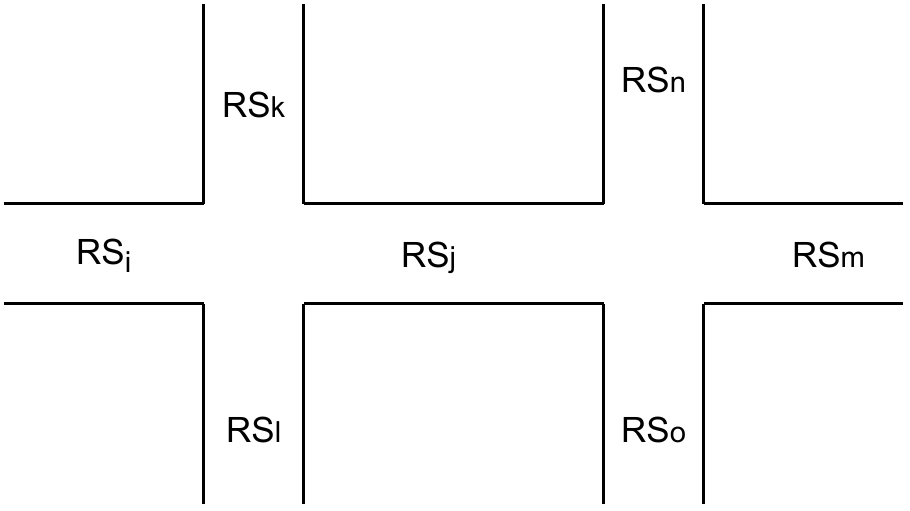}
\caption{Mobility modeling for selected road segments at  intersections}
\label{fig:MobilityModelling}
\end{figure}

\begin{equation}
\begin{split}
\forall{i_1 \in \{1,\ldots,\mathbb{I}\}; i_2 \in \{1,\ldots,\mathbb{I}\}; i_1 \neq i_2}\\
P_{t_1,t_2}(i)\{VS_{n+1}=RS_j|VS_{n}=RS_i \} \\
=  \frac{\#(RS_i to RS_j)}{\#(RS_i)}
\end{split}
\label{eq:CCP}
\end{equation}

where $[t_1,t_2]$ is the time interval selected based on the traffic state of the selected road segments. If the traffic is in a free-flow state, the time is chosen to be a 5-minute interval, and if there is queuing and the road is congested, the time interval is taken as 10 minutes. These values are inferred from our experiment of deriving microscopic probabilities for each vehicle at selected Dublin intersections, using the SUMO simulator. We use real vehicle density data from the Transport Infrastructure Ireland Traffic Data website\footnote{https://trafficdata.tii.ie/publicmultinodemap.asp}. We use vehicular counts to calibrate the traffic state of the selected road segments, thereby making the microscopic model more realistic.

\subsection{Objective Function}

We use node and link utilization cost as a measure to analyze the quality of placement of tasks on mobile vehicle nodes from the point of view of resource utilization. We aim to minimize the node and link utilization cost which is defined as:

\begin{enumerate}
    \item \textbf{Obj1} Node Utilization Cost: is defined as the ratio of total used computational capacity to the total available computational capacity for the service placement. The ratio is weighted by the CCP of the node. This considers mobility of nodes rather than placing tasks on nodes with high processing capacity but with very low CCP to stay with the other vehicles in the cluster. The node utilization cost is given as:
    \begin{equation}
\footnotesize{
\begin{aligned}
  \mbox{NodeCost} (i_1) = \sum_{\forall{i_1, i_2; i_1 \neq i_2}}\\ (1 - P_{(t_1,t_2)}(i)).(D_{pj,k}/C_{k}(i)).M(p,j,i)
\end{aligned}
}
\label{eq:NodeCapacityCost}
\end{equation}
where $P_{(t_1,t_2)}(i)$ is the CCP of the node with the TI $s_{pj}$ placed on it. As the CCP of a node increases, the cost of placing the TI on that node decreases. 
    \item \textbf{Obj2} Link Utilization Cost: The link utilization cost is defined as the ratio of total used link capacity to the total available link capacity. This ratio is weighted by the CCP of the link. The CCP of the link defines the joint probability of two nodes to stay together during the time window $[t_1,t_2]$. The link utilization cost is given as:
    \begin{equation}
\footnotesize{
\begin{aligned}
  \mbox{LinkCost} (i_1,i_2) =
\sum_{\forall{i_1, i_2; i_1 \neq i_2}} (1 - P_{(t_1,t_2)}(i_1,i_2)).\\ (F(p_1,p_2)/\sum path[B(i_1,i_2)])\Big(M(p,j_1,i_1).M(p,j_2,i_2)\Big)
\end{aligned}
}
\label{eq:LinkCost}
\end{equation}
where $P_{(t_1,t_2)(i_1,i_2)}$ is the joint CCP of two nodes to stay together and $path[B(i_1,i_2)]$ is a list of available bandwidth on the path between two nodes $i_1$ and $i_2$ with placed TIs. The total bandwidth is summed over the path between nodes $i_1$ and $i_2$.   

\item \textbf{Obj3} Chain length/hop count: The number of hops for the distributed service can have a significant impact on both communication overhead and service reliability. To keep the hops between two placed TIs to the minimum, we make sure to minimize the length of path between two nodes $i_1$ and $i_2$ with placed TIs. The network distance between two placed TIs is given as:

\begin{equation}\centering
\footnotesize{
\begin{aligned}
\forall{i_1 \in \{1,\ldots,\mathbb{I'}_{(pj)((p+1)j)}\}; i_2 \in \{1,\ldots,\mathbb{I'}_{(pj)((p+1)j)}\}; i_1 \neq i_2}\\
H(i_1,i_2) = \sum_{\forall{p_1,p_2}} M(p,j,i_1),M((p+1),j,i_2)  \\ len(path[B(i_1,i_2)]), \\
\end{aligned}
}
\label{eq:AdjacencyConstraints}
\end{equation}
where $H(i_1,i_2)$ is the hop count between two nodes $i_1$, $i_2$ with placed TIs. 
\end{enumerate}

The multi-objective function aims to minimize Obj1, Obj2, and Obj3. Each objective is weighted by $\lambda_1$, $\lambda_2$ and $\lambda_3$ such that $\lambda_1$ + $\lambda_2$ + $\lambda_3$ totals to 1 and each objective is weighted equally (but this can be changed to reflect operational requirements). The objective function is given as:

\begin{equation}
\footnotesize{
  \min\sum_{\forall{i_1, i_2; i_1 \neq i_2}}  \lambda_1 H(i_1,i_2) + \lambda_2 \mbox{LinkCost} (i_1,i_2) + \lambda_3 \mbox{NodeCost} (i_1) \\
}
\label{eq:ObjectiveFunction}
\end{equation}
where $H(i_1,i_2)$ is the hop count between two nodes $i_1$, $i_2$ with placed TIs. 

 \section{Heuristic-based Solution}

\begin{algorithm*}
\caption{Service Placement}\label{alg:Service Placement}
    \hspace*{\algorithmicindent} \textbf{Input:} \text{LG Linear type graph, Vehicle cluster graph} \\
    \hspace*{\algorithmicindent} \textbf{Input:} \text{$(UL_{Type_{i+1}},LL_{Type_{i+1}})$: Upper and lower limit for number of Type i+1 TI} \\
    \hspace*{\algorithmicindent} \textbf{Output:} \text{Successful/Unsuccessful service placement} 
\begin{algorithmic}[1]
\Procedure{ServiceMapping}{$LG,VC$}\Comment{The g.c.d. of a and b}
\While{$Type_1$}  \Comment{For all Type 1 instances}
    \For{\texttt{($Type_1,CN$)} $\in$ \text{TI\_pairs}}
        \For{\text{i} $\in$ \text{TI\_pairs($Type_1,CN$)}} \Comment{Ensures placement of full chain for each $Type_{i+1}$ instance}
            \State \text{TI\_placement(i,VC,$(UL_{Type_{i+1}},LL_{Type_{i+1}})$)}
            \If {\text{placement is successful}}
               \State {\text{Success}}
            \EndIf    
        \EndFor 
    \EndFor
\EndWhile
\EndProcedure
\Procedure{TI\_placement}{i,VC,($(UL_{Type_{i+1}},LL_{Type_{i+1}})$)}
\While{(i)}  \Comment{for all available $Type_1$ TCI} 
   \For{($Type_i$,$Type_{i+1}$) $\in$ i}  
       \State \text{$node_1 \gets$ location of $Type_{i}$}
             \If{\text{$Type_{i+1}$ instances exist on VC}} \Comment{To re-use instances}
             \State \text{$node_2$ $\gets$ List of location of $Type_{i+1}$}
             \For{j $in$ $node_2$}
                  \If{\text{resource at j  $\geq$ resource required for $Type_{i+1}$}}
                      \State \text{$pathtonode_2$ $\gets$ Get path from $Type_i$ to $Type_{i+1}$}
                  \EndIf  
                  \State \text{$sortedpath$ $\gets$ sorts path based on path length} 
                  \For{\text{k in $sortedpath$}}
                      \If{\text{$required_{datarate}$ $\leq$ $min(k\_path\_datarate(i,j))$}}
                          \State \text{place $Type_{i+1}$ on $node_2$}
                          \State{\Return{TI $Type_{i+1}$ placed}}
                          \State{\text{Break}} 
                     \EndIf      
                  \EndFor
            \EndFor
            \Else
              \If{\text{length($node_2$) $\geq$ $UL_{Type_{i+1}}$}}
                \State{\Return{Not enough resources on vehicle cluster}}
              \Else

                 \State{$CN_{location}$ $ \gets$ location of CN}
                 \State{paths $\gets$ weighted path from $Type_{i+1}$ to CN}
                 \State{$sorted\_paths$ $\gets$ sort paths from highest to lowest path weight}
                 \For{i in $sorted\_paths$}
                    \If{\text{$required_{datarate}$ $\leq$ $min(i\_path\_datarate(x,y))$}}
                    \While{nodes available in i}
                       \State{v $\gets$ next node on i}
                       \If{resource on v $\geq$ resource required by $Type_{i+1}$}
                          \State{Place $Type_{i+1}$ on v}
                          \State{\text{break}}
                        \EndIf
                        \If{if no node available on path($Type_{i}$,CN)}  
                             \State{\Return{{Unable to place TI on cluster}}}
                         \EndIf
                 \EndWhile 
                 \EndIf
                 \EndFor
              \EndIf
           \EndIf
    \EndFor
\EndWhile
\EndProcedure
\end{algorithmic}
\end{algorithm*} 

Due to the nature of vehicular networking, it is required to scale services and find efficient service placement in a very short time, compared to the time required to solve the full integer-linear program (ILP). Consequently, we propose a node selection and service placement \emph{heuristic} solution. We first select a vehicular cluster based on the vehicle's mobility behavior by using the principles of community detection. 

The mobility of vehicle nodes is constrained by the underlying road topology. We model the available vehicle cluster as a graph with their joint CCP as the edge weight for each edge, depicting how probable are two nodes to stay together in the next time segment. The use of community detection using mobility behavior helps in identifying the most connected nodes that play a crucial role in the forwarding of data flows. The identification of communities helps in reducing service reconfiguration, especially for a distributed service with data-dependent TIs. 

\subsection{Node Selection} We use community detection, which is the process of discovering cohesive groups or clusters in a network, to determine nodes that have better connectivity between them than the rest of the network. Using community detection algorithms, we partition the network graph into communities and, the biggest community is chosen for service placement. We consider the CCP as the edge weight in the network such that nodes that have a higher probability to stay connected are chosen for service placement. Due to the data-dependency between TIs in the service model, all TI's are promoted to be placed in the same community of nodes. 

We analyze two community detection algorithms for partitioning the vehicle cluster for service deployment. We first use a modularity score-based Louvain algorithm \cite{2008} that initially starts with $|V|$ communities where each node is considered to be a community in the first iteration. Modularity is defined as the density of edges inside the community with respect to edges outside the community. In each iteration, every node is moved to its neighboring community and the gain in modularity is calculated. If the gain is positive, the node does not return to its previous community. The iterations of the heuristic stop when the modularity gain, between any two iterations, does not exceed a specified threshold value. The algorithm has the complexity of $O(V log V)$ where V is the number of nodes in the graph. In our experiment, the modularity obtained in a graph of 30 nodes was 0.1428. 

We also considered the hierarchical clustering-based Girvan and Newman algorithm \cite{PhysRevE.70.056131} which derives a community tree or a dendrogram with a specified depth \cite{8588297}. The connectivity of a community increases as the depth of the derived dendrogram increases. The method first removes the edge with the highest edge betweenness centrality. The edge betweenness centrality is the sum of the fraction of the shortest paths that cross the edge. Each iteration splits every existing community into two new communities. The disconnected sub-graphs undergo the same procedure until the entire graph is split into isolated nodes. The complexity of the algorithm is $O(E^2V)$, where E is the edges of the graph and V represents the nodes. The modularity score for the same graph using this method is 0.00186. We, therefore, prefer the Louvain method as it results in a higher modularity score, which is more useful in this context, and Louvain's computational complexity is also lower.

\begin{figure}[htbp]
\begin{subfigure}{0.5\linewidth}
  \includegraphics[width=\linewidth,height=3cm,keepaspectratio]{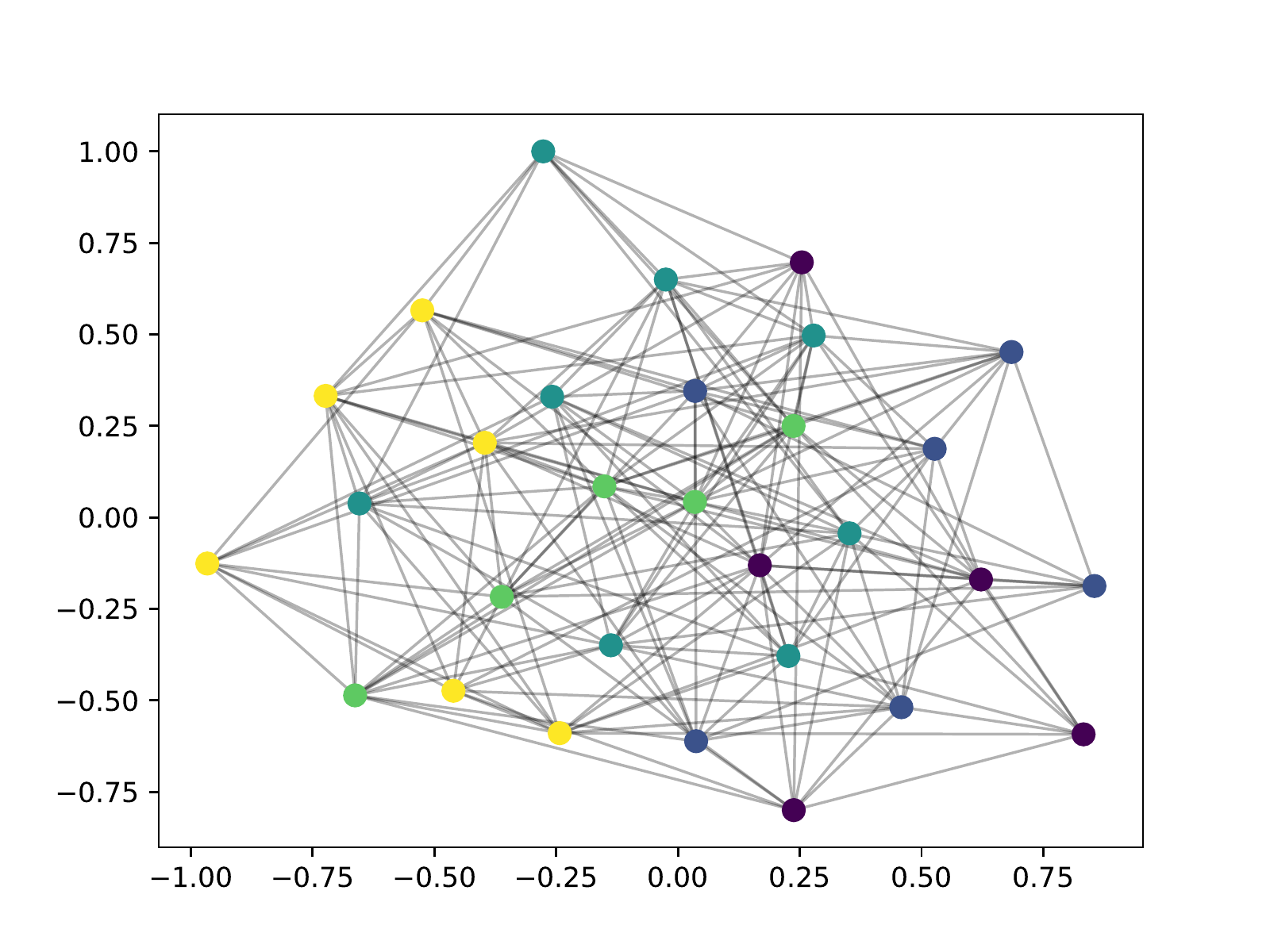}
  \caption{Louvain method}
  \label{fig:PenaltyCost-ResourceConstrained}
\end{subfigure}\hfil 
\begin{subfigure}{0.5\linewidth}
  \includegraphics[width=\linewidth,height=3cm,keepaspectratio]{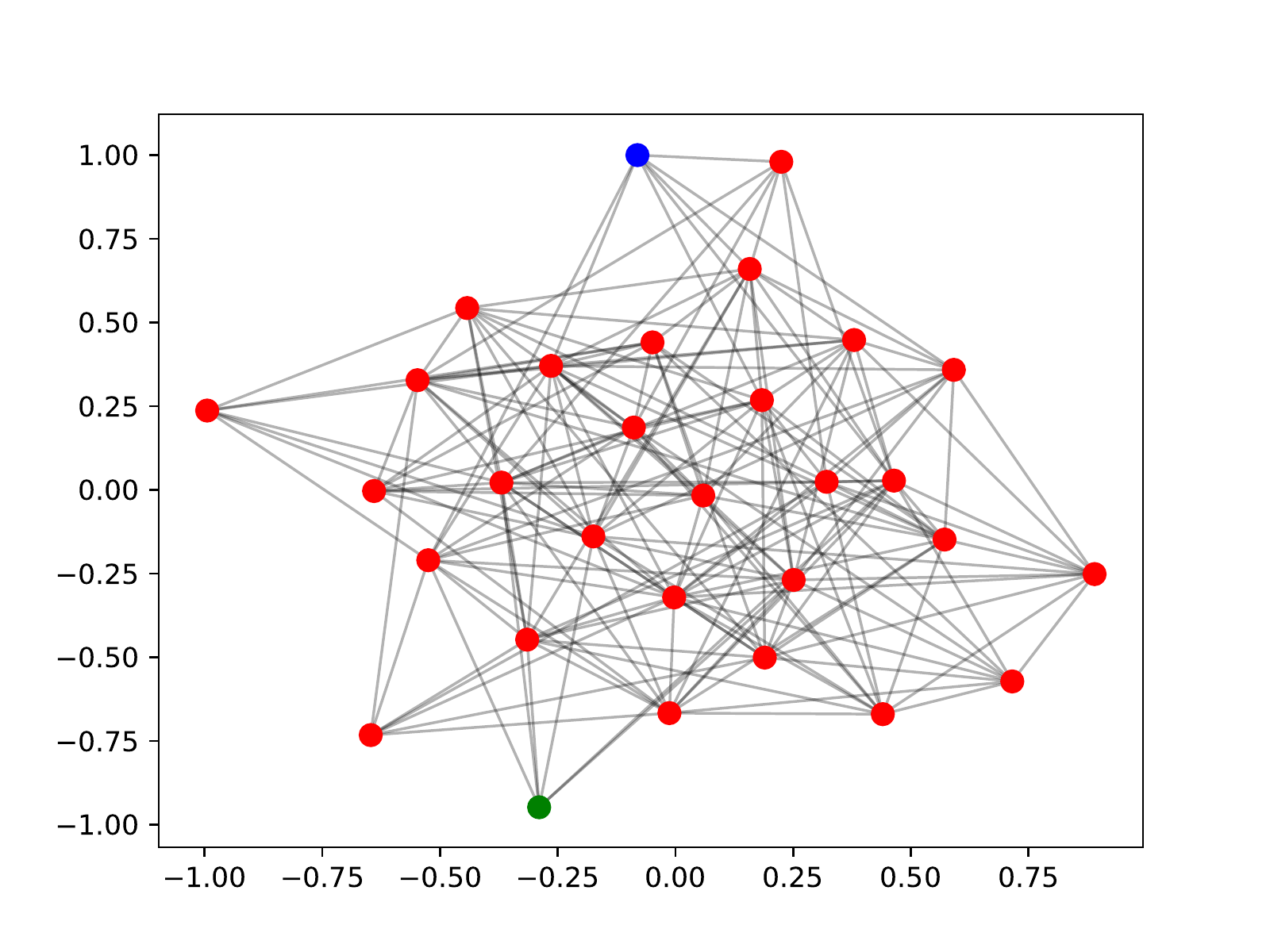}
  \caption{Girvan and Newman method}
  \label{fig:non-weighted}
\end{subfigure}\hfil 
\caption{Using two different method for community/cluster detection}
\label{fig:weighted}
\end{figure}

\subsection{Service placement heuristic}

After the vehicle cluster is selected based on the CCP of nodes using the community detection method, we then place the TIs by utilizing a graph-based heuristic. We first give as input, 1) the selected vehicular cluster which is the strongest detected community, 2) the linear type graph to be placed and 3) the upper $(\textrm{UL}_{\textrm{Type}_{i}})$ and lower limit $(\textrm{LL}_{\textrm{Type}_{i}})$ for the number of TIs of each type to be placed. The LL for all the tasks is 1 as we want to make sure at least one TI of each type is placed. The UL for each TI is equal to the number of video sources or Type 1 TIs, ensuring each stream gets one processing TI, in case the available processing capacity at individual nodes is very low. 

\begin{figure}[htbp]\centering
\includegraphics[width=\linewidth,height=3cm]{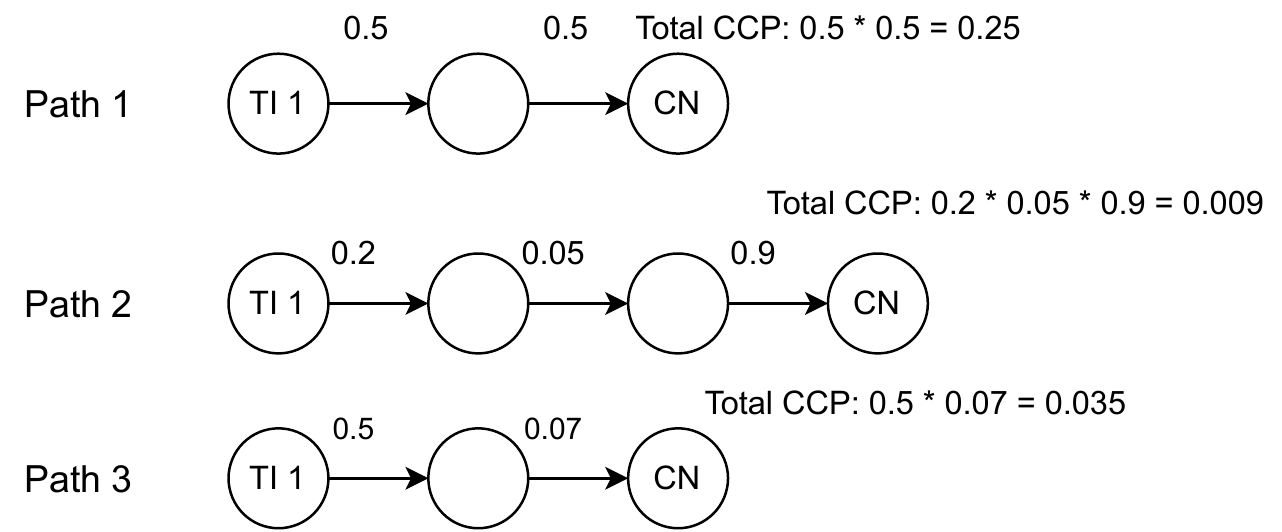}
\caption{Joint CCP-based path selection for service placement}
\label{fig:Pathlen}
\end{figure}

We modify a heuristic algorithm inspired by the work of \cite{nem.1963}, where VNFs are placed along the shortest path with the smallest bottleneck value. Instead of placing TIs on the shortest path, we consider the joint CCP of the path from a source TI (of Type 1) to the CN, as depicted in Fig.\ref{fig:Pathlen}. We then place TIs along the path with the highest joint CCP. As we intend to place a long chain of TIs along this path, choosing a longer chain increases the possibility of placing most TIs on the path to the CN. The heuristic may also randomly choose the shortest path, in terms of hop count, if the combined CCP of the path is the highest. From the three example paths shown in Fig.\ref{fig:Pathlen}, the heuristic will choose path 1 as it has the highest combined CCP, even though it is shorter than path 2. Choosing path 2 will result in placing TI on two nodes linked by very low CCP, 0.05 in this case. Path 3 is the same length as Path 1 but has lower joint CCP in comparison. 

The heuristic is described as Algorithm 1. Similar to  \cite{nem.1963}, we place TIs in a pairwise way, as the service model has dependent TIs. In our model, every TI of any type has a common endpoint as the CN. In line 3, we iterate over all TI pairs from Type 1 to the CN. In line 5, each TI pair is sent to the TI\_PLACEMENT function along with the UL and LL for the TI. In line 11, the location of the Type 1 instance is detected. On line 12 it is checked if the next TI, of type $Type_{i+1}$, exists on the vehicle cluster. If it exists, say at node j, and the resource capacity at node j meets the capacity constraint for TI $Type_{i+1}$, all the paths from $Type_i$ to $Type_{i+1}$ are stored in the list $sortednode_2$. In line 18, all the available paths are iterated over and the bottleneck edge capacity for each path is compared to the required available capacity between the two TIs. If the constraint is met, $Type_{i+1}$ TI is reused for the flow. If $Type_{i+1}$ does not exist on the cluster, it is checked if the upper limit for the TI type is met (on line 24). 

If the upper limit is not reached, all the paths are explored from Type$_i$ to CN of the cluster. All the paths are sorted based on the path weight, which in our case is the total CCP of the path. In line 31, all the paths are iterated over, and the bandwidth capacity requirement is checked for the path. If the bandwidth requirement is met and the resource capacity requirement for the node is met, then Type$_{i+1}$ is placed on the node v. If there are no more available nodes on the path to the CN, a failed placement is registered. Thus, this approach aims to send the collected data to the CN and tries to place processing TIs in-network when possible.

\begin{table}[]
\centering
\begin{tabular}{|c|c|}
\hline
\textbf{\begin{tabular}[c]{@{}c@{}}Number of \\ Type 1 instances\end{tabular}} & \textbf{\begin{tabular}[c]{@{}c@{}}Optimality \\ gap (\%)\end{tabular}} \\ \hline
1 & 0      \\ \hline
2 & 0      \\ \hline
3 & 11.03  \\ \hline
4 & 13.05  \\ \hline
5 & 10.997 \\ \hline
6 & 17.83  \\ \hline
\end{tabular}
\caption{Optimality gap percentage for the link utilization cost for different number of Type 1 TIs}
\label{tab:Optimalitygap}
\end{table}

\begin{figure*}[htb]
    \centering 
\begin{subfigure}{0.3\textwidth}
  \includegraphics[width=\linewidth,height=5cm]{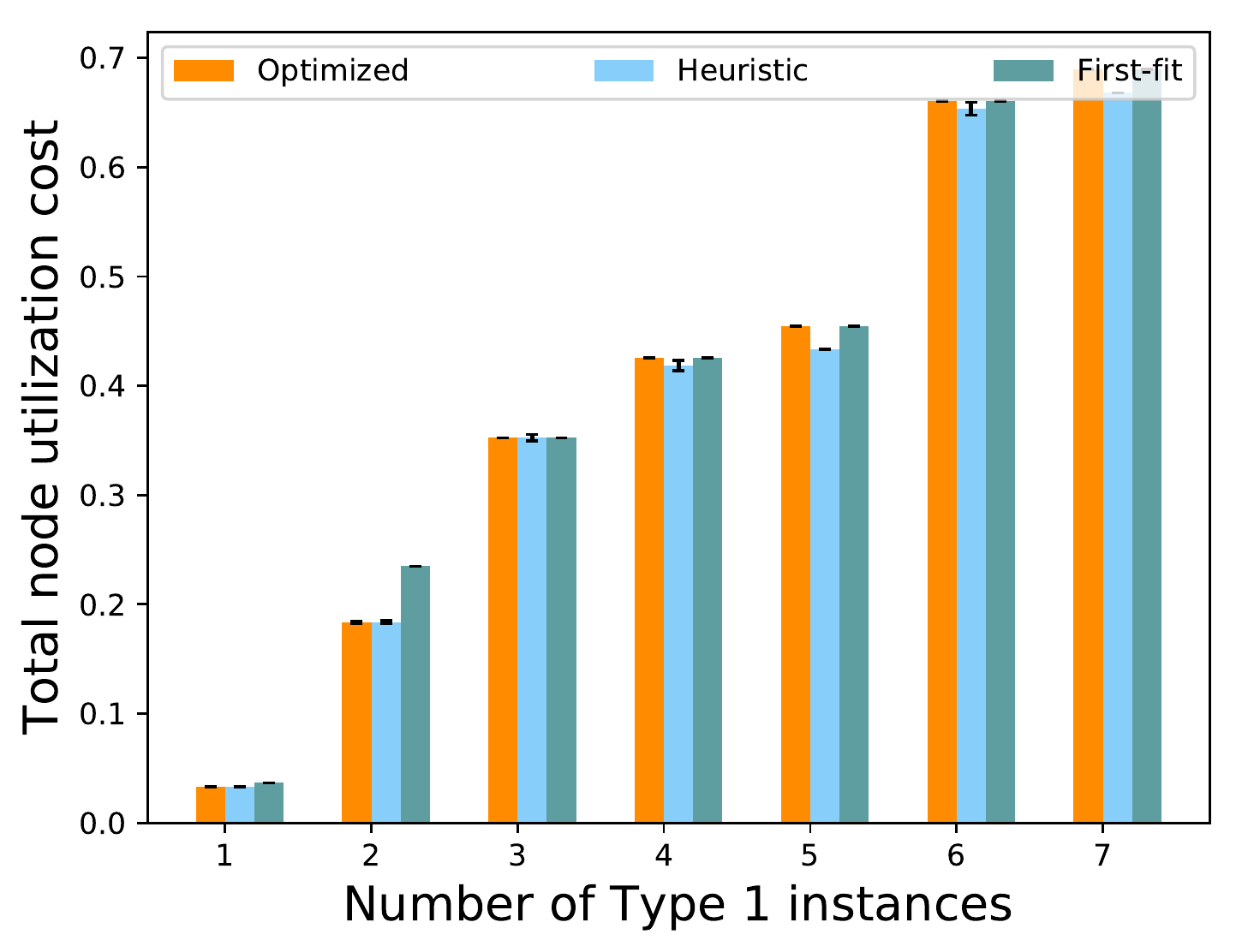}\vspace{-0.75em}
  \caption{\textbf{3 task chain: Total node utilization cost}}
  \label{fig:NodeCostCase-a}
\end{subfigure}\hfil 
\begin{subfigure}{0.3\textwidth}
  \includegraphics[width=\linewidth,height=5cm]{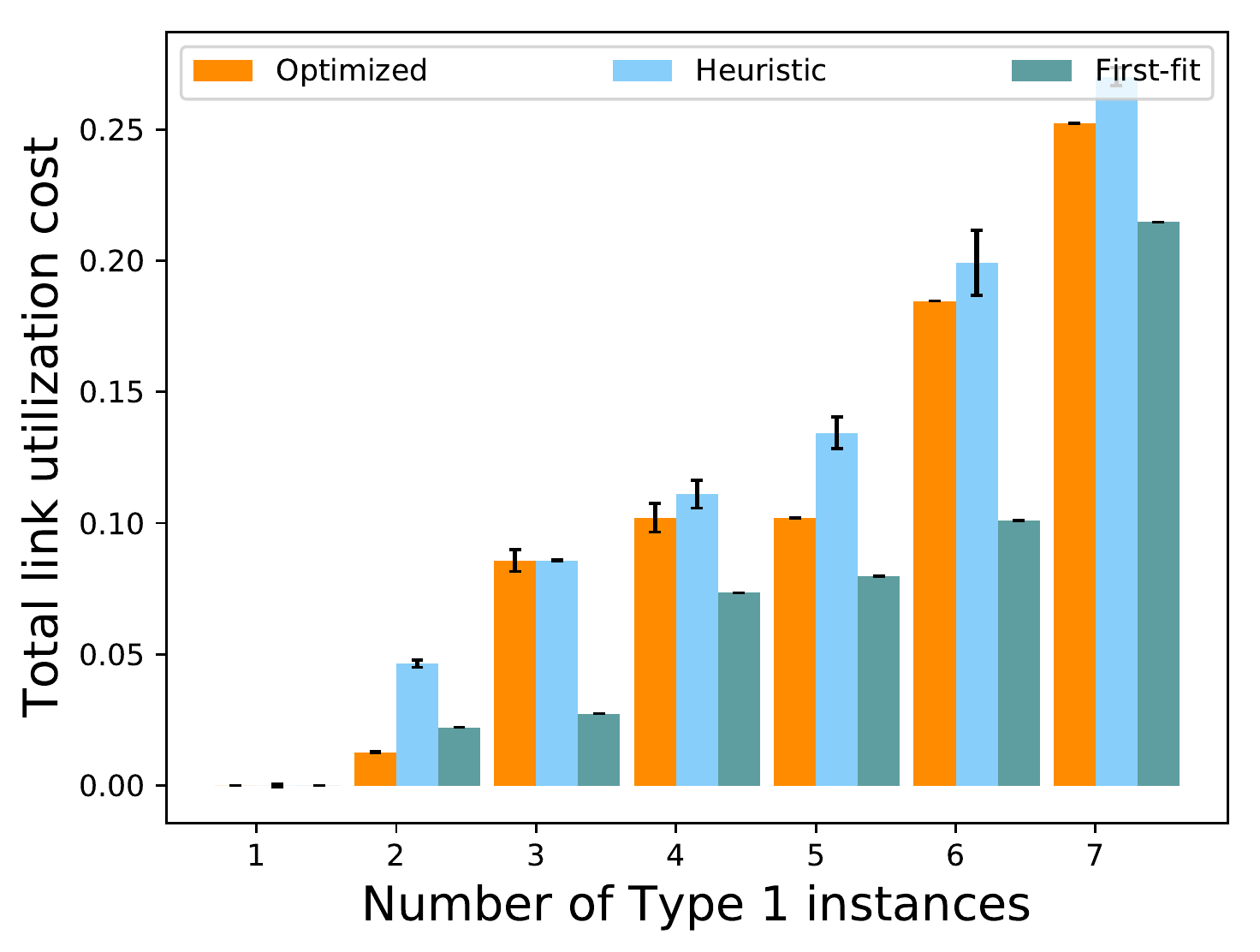}\vspace{-0.75em}
  \caption{\textbf{3 task chain: Total link utilization cost}}
  \label{fig:LinkCostCase-a}
\end{subfigure}\hfil 
\begin{subfigure}{0.3\textwidth}
  \includegraphics[width=\linewidth,height=5cm]{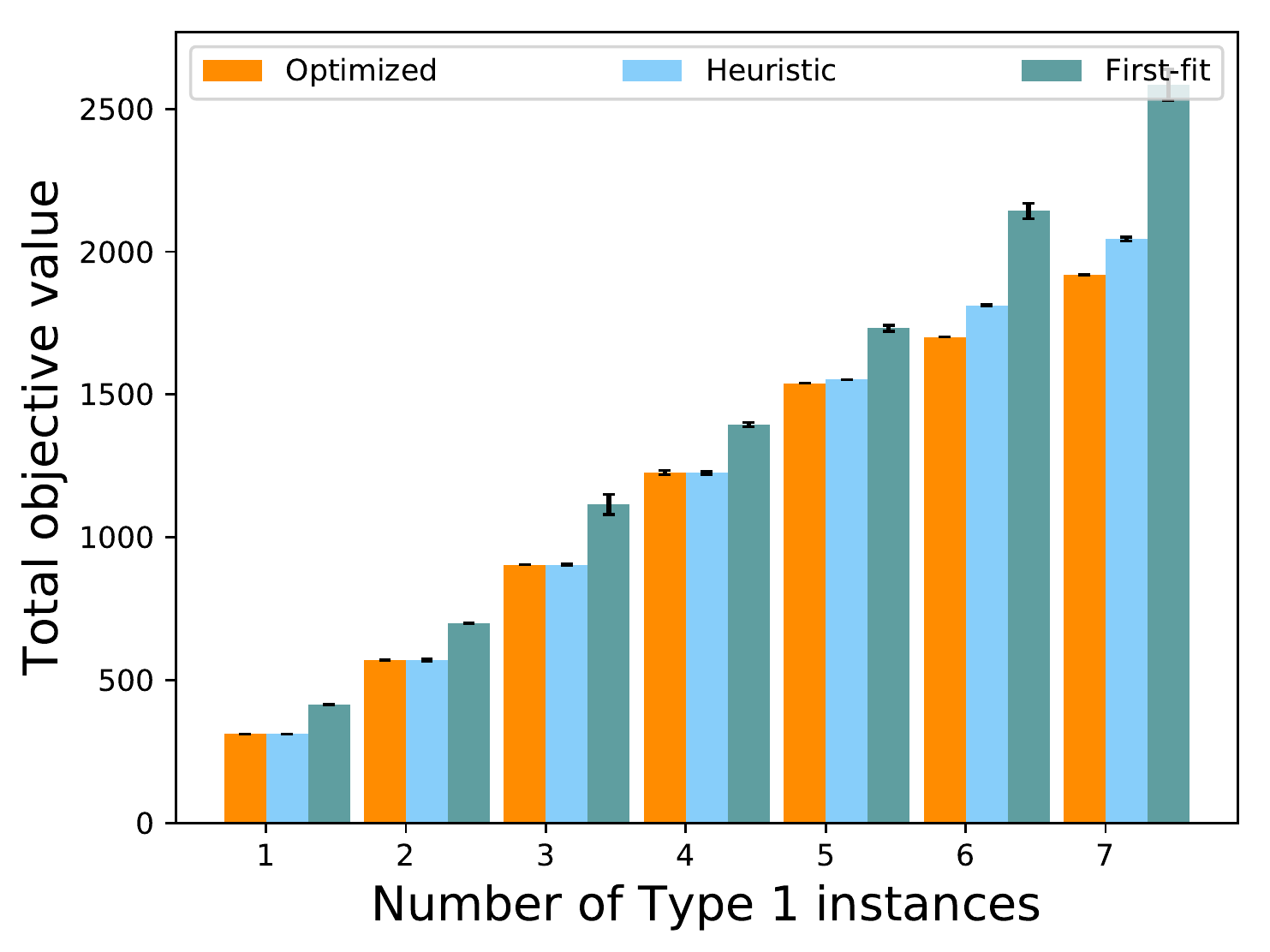}\vspace{-0.75em}
  \caption{\textbf{3 task chain: Comparison of the total objective value}}
  \label{fig:totalobjCase-a}
\end{subfigure}

\medskip
\begin{subfigure}{0.3\textwidth}
  \includegraphics[width=\linewidth,height=5cm]{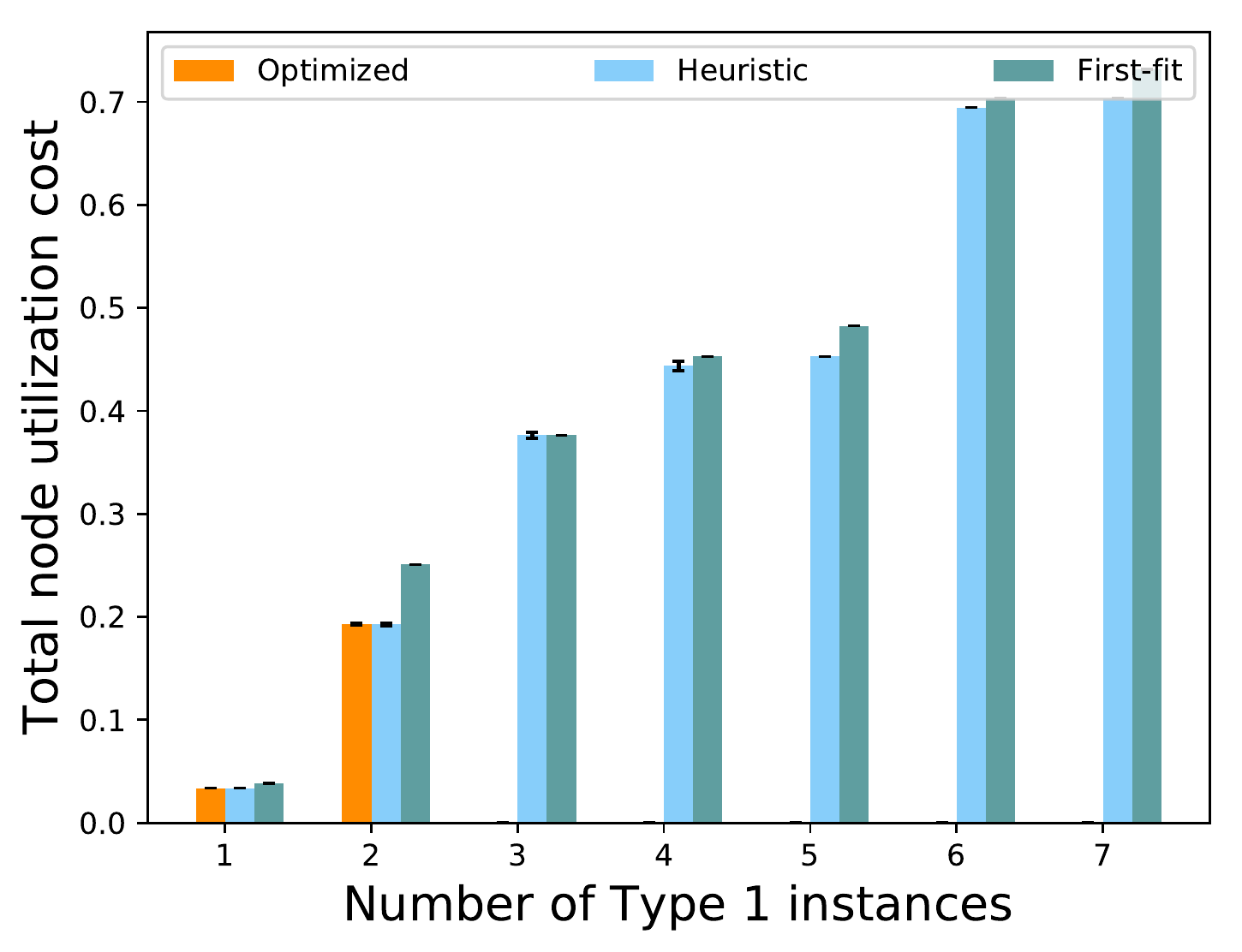}\vspace{-0.75em}
  \caption{\textbf{4 task chain: Total node utilization cost}}
  \label{fig:NodeCostCase-b}
\end{subfigure}\hfil 
\begin{subfigure}{0.3\textwidth}
  \includegraphics[width=\linewidth,height=5cm]{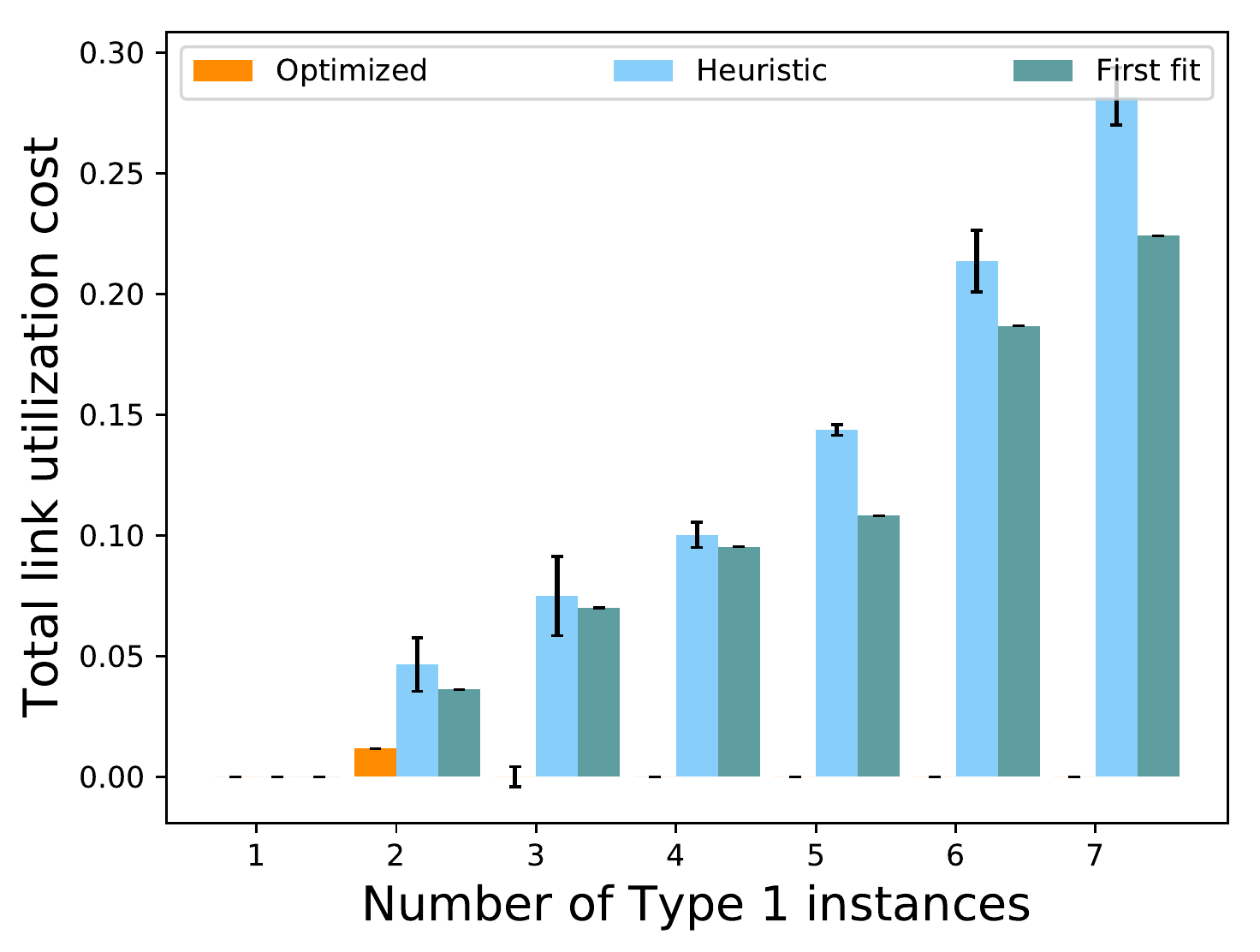}\vspace{-0.75em}
  \caption{\textbf{4 task chain: Total link utilization cost}}
  \label{fig:LinkCostCase-b}
\end{subfigure}\hfil 
\begin{subfigure}{0.3\textwidth}
  \includegraphics[width=\linewidth,height=5cm]{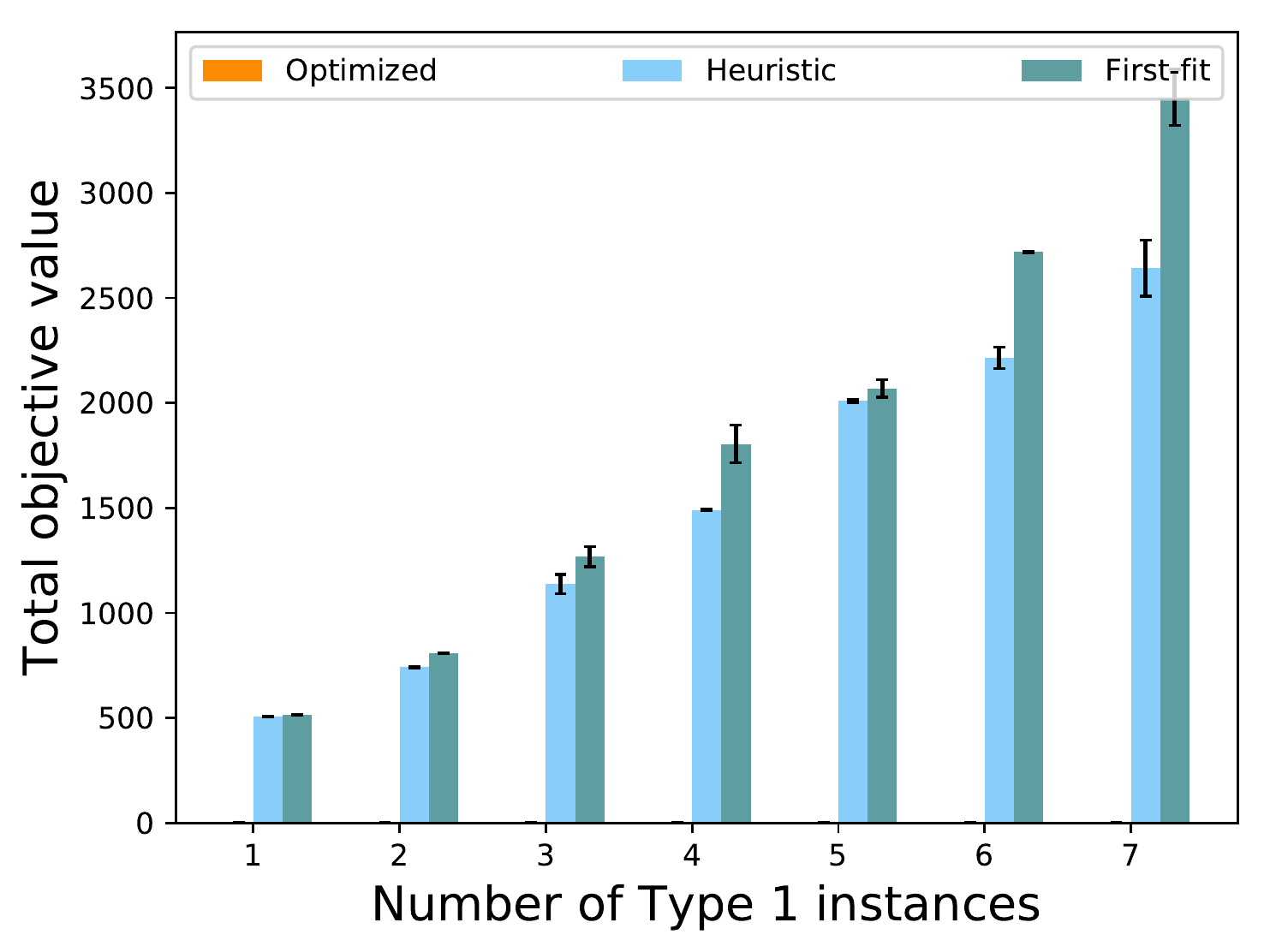}\vspace{-0.75em}
  \caption{\textbf{4 task chain: Comparison of the total objective value}}
  \label{fig:totalobjCase-b}
\end{subfigure}
\caption{Total node utilization cost, total link utilization cost and total objective value for task chain of lengths 3 and 4}
\end{figure*}

\section{Evaluation}

In this section, we evaluate the performance of the Mixed Integer-linear Program (MIP), the proposed heuristic, and the first-fit approach through resource utilization metrics for bandwidth (link) and processing (node). We also compare the MIP, the first-fit, and the heuristic for the total optimal value. We then evaluate the average chain length and the total instances that the heuristic scales to analyze the performance of the heuristic in terms of not overprovisioning resources. We then use total response time as a Quality-of-Service measure and compare our placement approach to a static mobile-edge computing approach, which does not use task replication in the form of multiple instances of the same task. We also evaluate the performance of the selected vehicular cluster using centrality measures. We use vehicular mobility on a Fog-computing-based simulator, to study the evolution of the selected cluster through its lifetime. 

We first use Gurobi, a standard MIP solver to solve the multiple objectives, constrained optimization problem. We find an optimal solution for placing both service types of different chain lengths on a selected vehicle cluster. We evaluate the solution for the node utilization cost, link utilization cost, and the total objective value for placing multiple applications of chain lengths 3 and 4. We vary the number of video instances from 1 to 7 to evaluate the scalability of the experiment. The applications are defined in Section~\ref{sec:System Model}, where the `data collection application' type is rich in data flow and has low processing requirements, whereas the `object detection' application type is more compute-intensive and has less bandwidth requirement. The applications are of variable chain length.  We use the first fit approach as the baseline approach. It sorts all the available paths from the data collecting TI to the CN and sorts them based on the highest available resource capacity. It then places TIs on all the available vehicle nodes on that path.  

We first place two applications of the first type on the selected vehicular cluster. For the case of the 3 task chain, our heuristic gives better node utilization cost as compared to both optimal and first-fit solutions, as shown in Fig.\ref{fig:NodeCostCase-a}. Our heuristic gives comparable link utilization cost in comparison to the optimal solution for 1-3 Type 1 TIs, but it becomes less efficient for a higher number of Type 1 TIs, as shown in Fig.\ref{fig:LinkCostCase-a}. This is due to prioritizing paths with higher CCP which may result in selecting longer routes between dependable TIs. 

For the total objective value, our heuristic performs as well as the optimal solution for the 1-5 Type 1 TIs, as shown in Fig.\ref{fig:totalobjCase-a}. For more number video TIs (Type 1), our heuristic under-performs when compared to the optimal solution. The optimality gap percentage has been summarised for the case in Table \ref{tab:Optimalitygap}. The worst-case optimality gap is 17.83\% for the case of 7 Type 1 TIs. The first-fit algorithm performs poorly for any number of Type 1 TI, irrespective of the scale. In terms of execution time, the ILP solves the problem in 1 second for 1 Type TIs, whereas the heuristic solution takes 0.1154 seconds. Each run is being performed on the same hardware, an Intel i7-6500U running at 2.50 GHz, not optimised for performance. For the case of 7 Type 1 TIs, the ILP solution takes 15.60 seconds whereas the heuristic solution takes 1.235 seconds. 

For the second case, we place both applications of chain length 3 and 4 TIs for both services. The ILP solver fails to give a solution in a reasonable time for applications with a longer chain length. We get a solution from the solver in seconds for 1-2 Type 1 TIs. But as the number of Type 1 TIs increases, the solver does not converge to a solution even after hours. We, therefore, compare our heuristic solution to the baseline approach. Our heuristic performs better than the baseline approach for any number of Type 1 TIs (from 1 to 7), as shown in Fig.\ref{fig:NodeCostCase-b}. The baseline approach outperforms the heuristics solution for the link utilization cost for the second case, as shown in Fig.\ref{fig:LinkCostCase-b}. This is due to choosing paths that have higher joint CCP, to increase the robustness of the service placement. This results in more bandwidth utilization as a tradeoff to selecting more robust paths. Our approach outperforms in minimizing the total objective value as compared to the first-fit approach, as depicted in Fig.\ref{fig:totalobjCase-b}. The baseline approach fails in minimizing the total objective for the higher number of Type 1 TIs.

\begin{figure*}[htb]
    \centering 
\medskip
\begin{subfigure}{0.3\textwidth}
  \includegraphics[width=\linewidth,height=5cm]{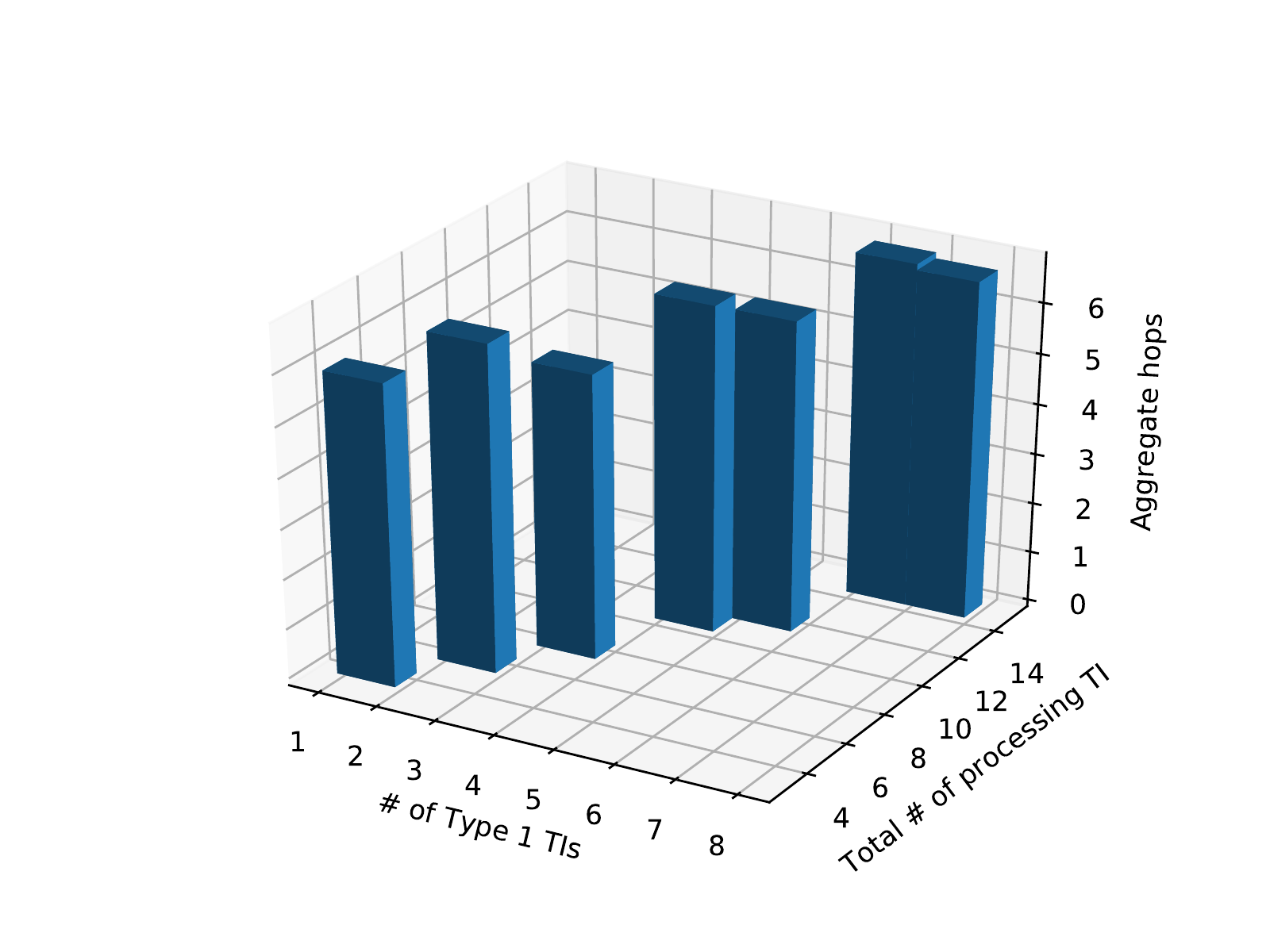}\vspace{-0.75em}
  \caption{\textbf{Comparison of total number of scaled, processing TIs and aggregate hops for the 3 TI chain placement corresponding to the number of Type 1 TIs}}
   \label{fig:hopcount}
\end{subfigure}
\begin{subfigure}{0.3\textwidth}
  \includegraphics[width=\linewidth,height=5cm]{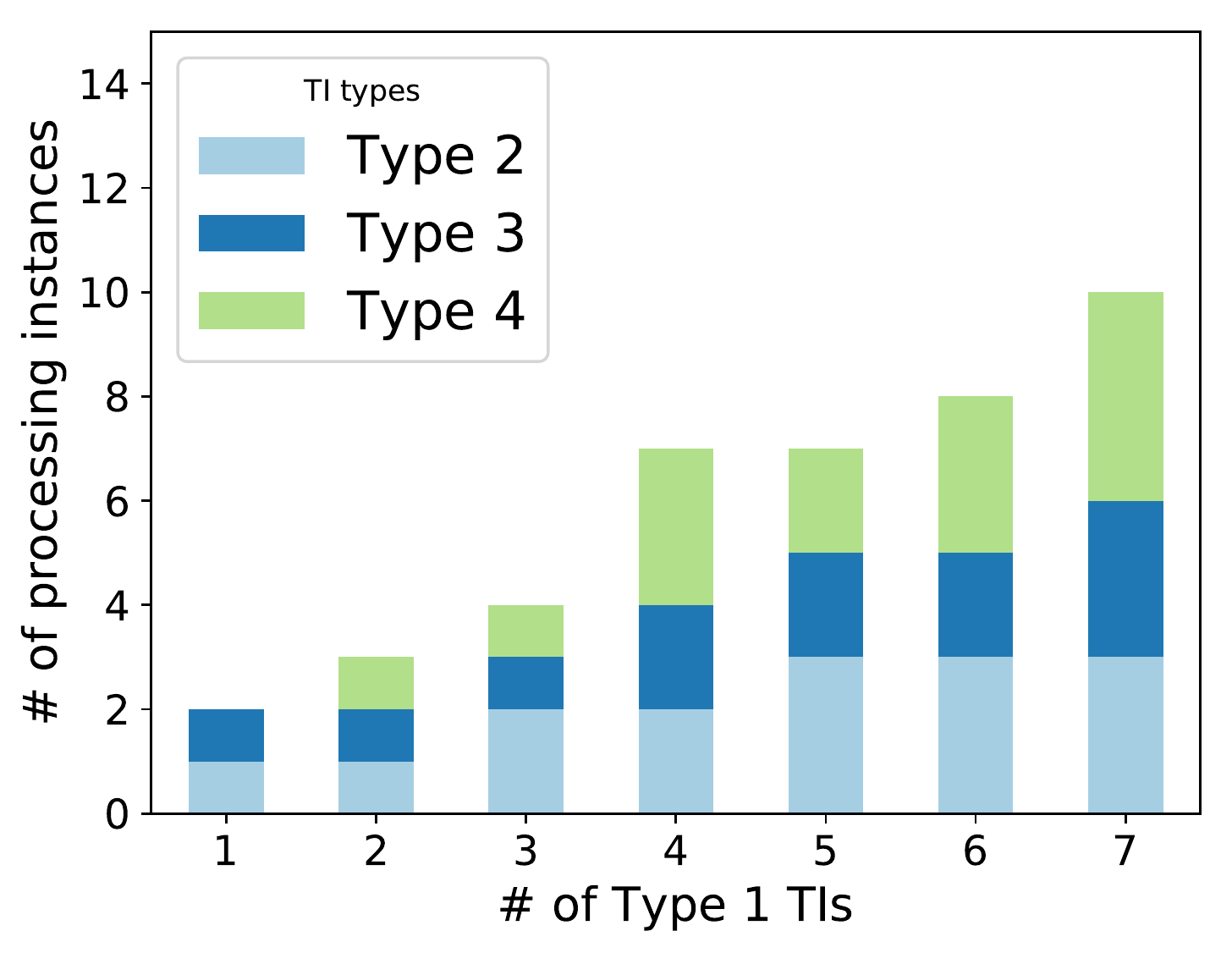}\vspace{-0.75em}
  \caption{\textbf{Minimum number of required processing TIs to meet the service demand corresponding to number of Type 1 TIs}}
  \label{fig:MinprocessingTI}
\end{subfigure}  
\begin{subfigure}{0.3\textwidth}
  \includegraphics[width=\linewidth,height=5cm]{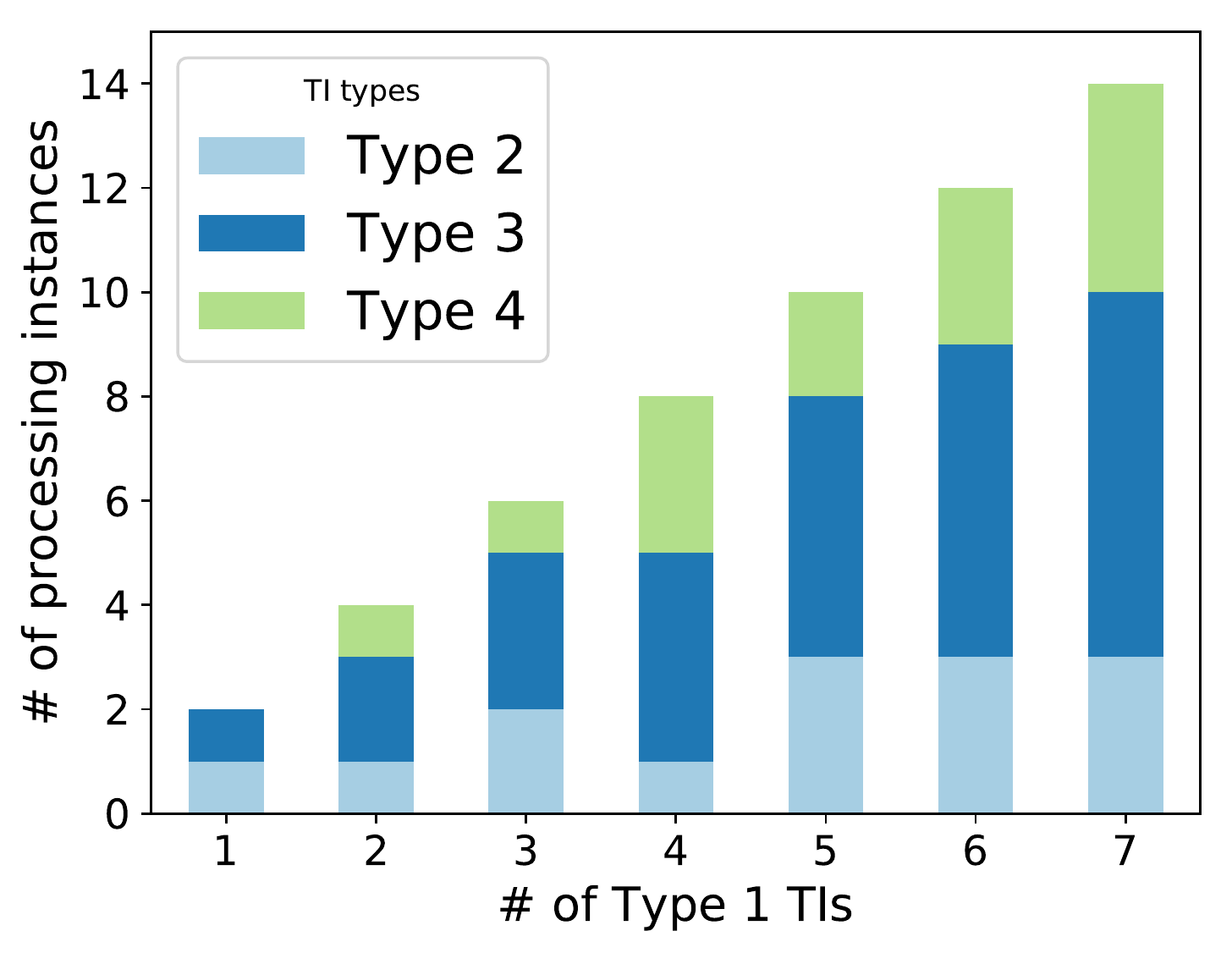}\vspace{-0.75em}
  \caption{\textbf{Total number of processing TIs scaled by our heuristic corresponding to the number of Type 1 TIs }}
  \label{fig:processingTI}
\end{subfigure}\hfil 

\medskip

\begin{subfigure}{0.3\textwidth}
  \includegraphics[width=\linewidth,height=5cm]{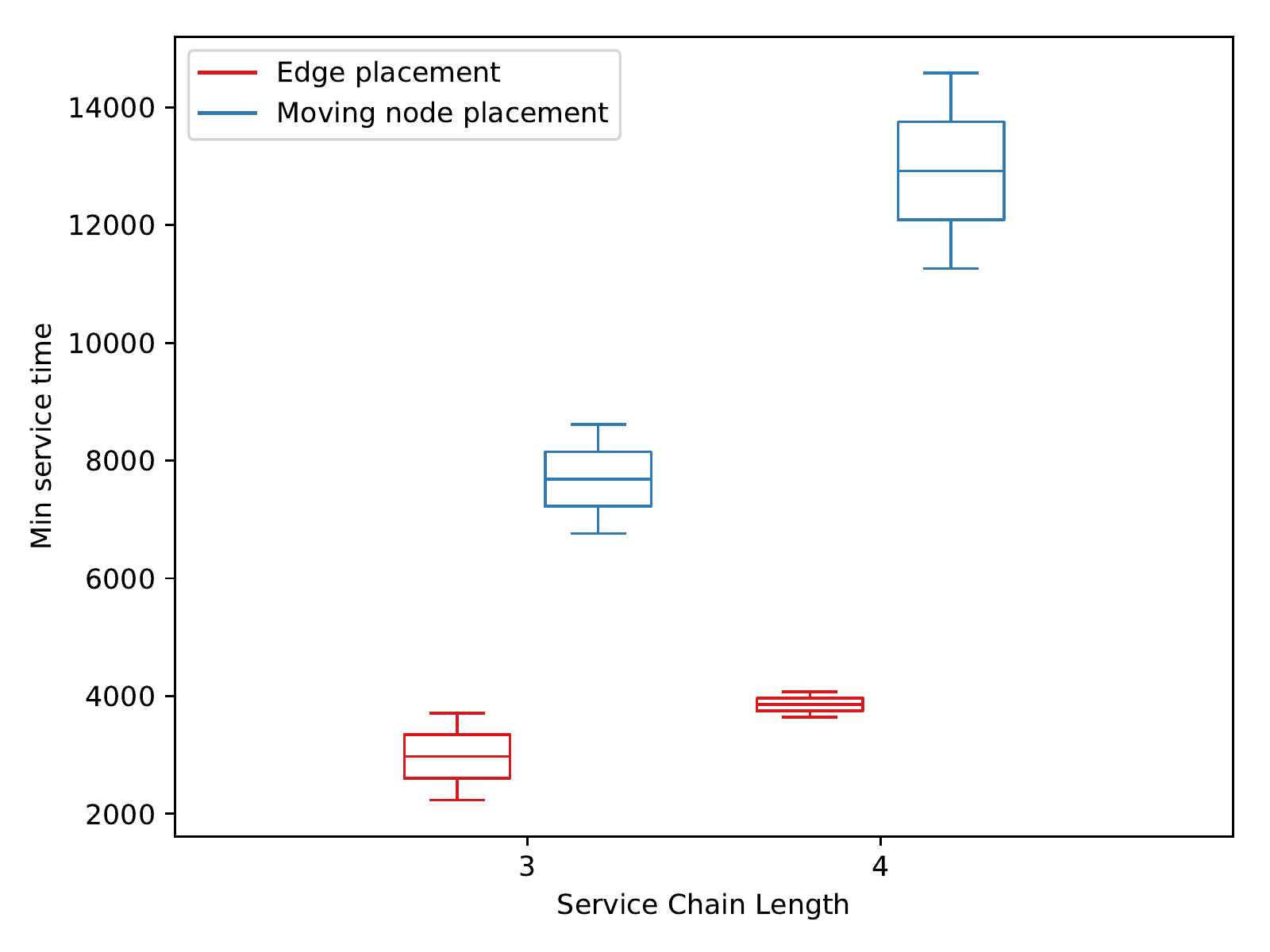}\vspace{-0.75em}
  \caption{\textbf{Minimum service time for different chain length}}
  \label{fig:Minservicetime}
\end{subfigure}\hfil 
\begin{subfigure}{0.3\textwidth}
  \includegraphics[width=\linewidth,height=5cm]{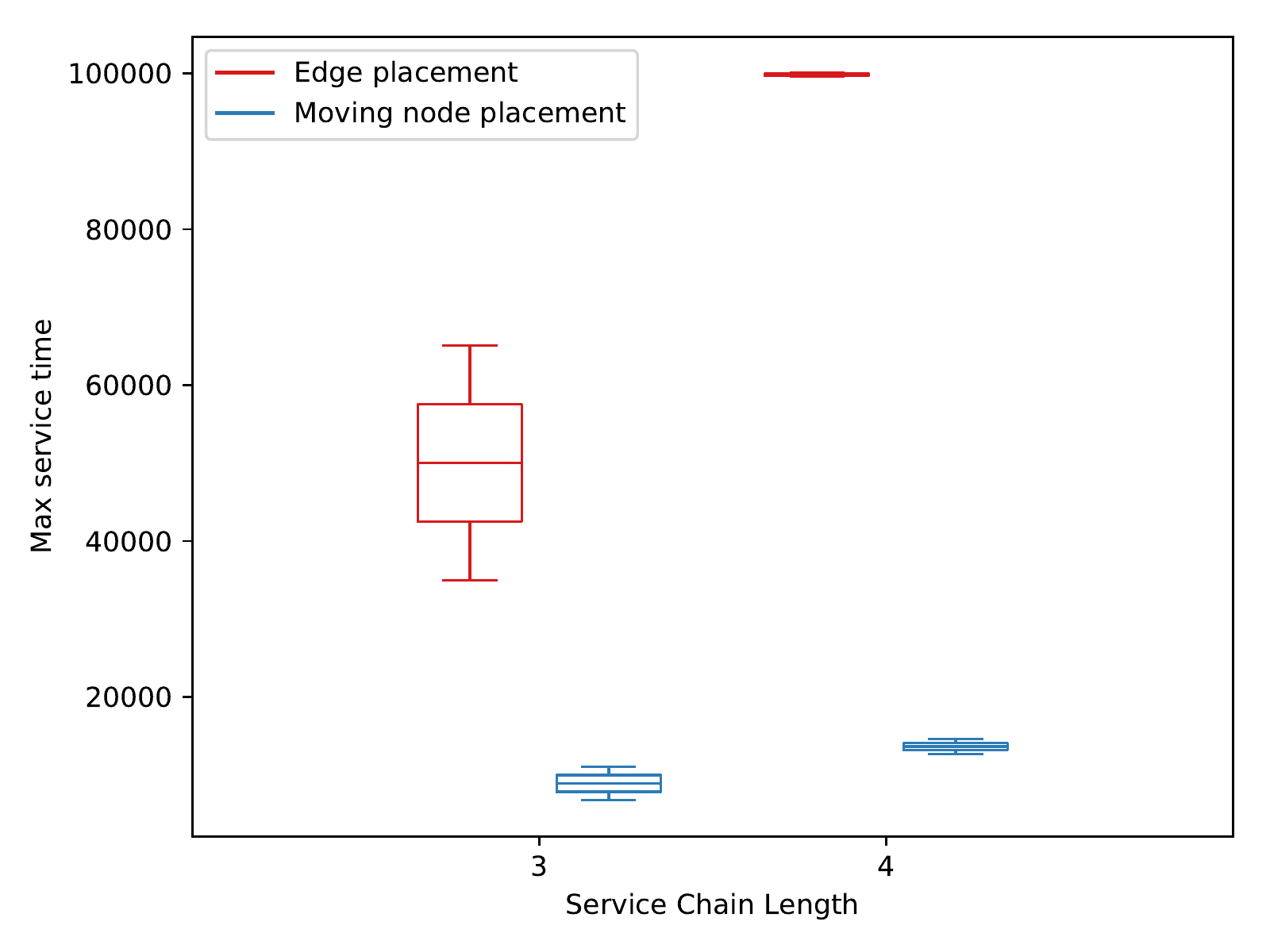}\vspace{-0.75em}
  \caption{\textbf{Maximum service time for different chain length}}
  \label{fig:Maxservicetime}
  
\end{subfigure}
\begin{subfigure}{0.3\textwidth}
    \includegraphics[width=\linewidth,height=5cm]{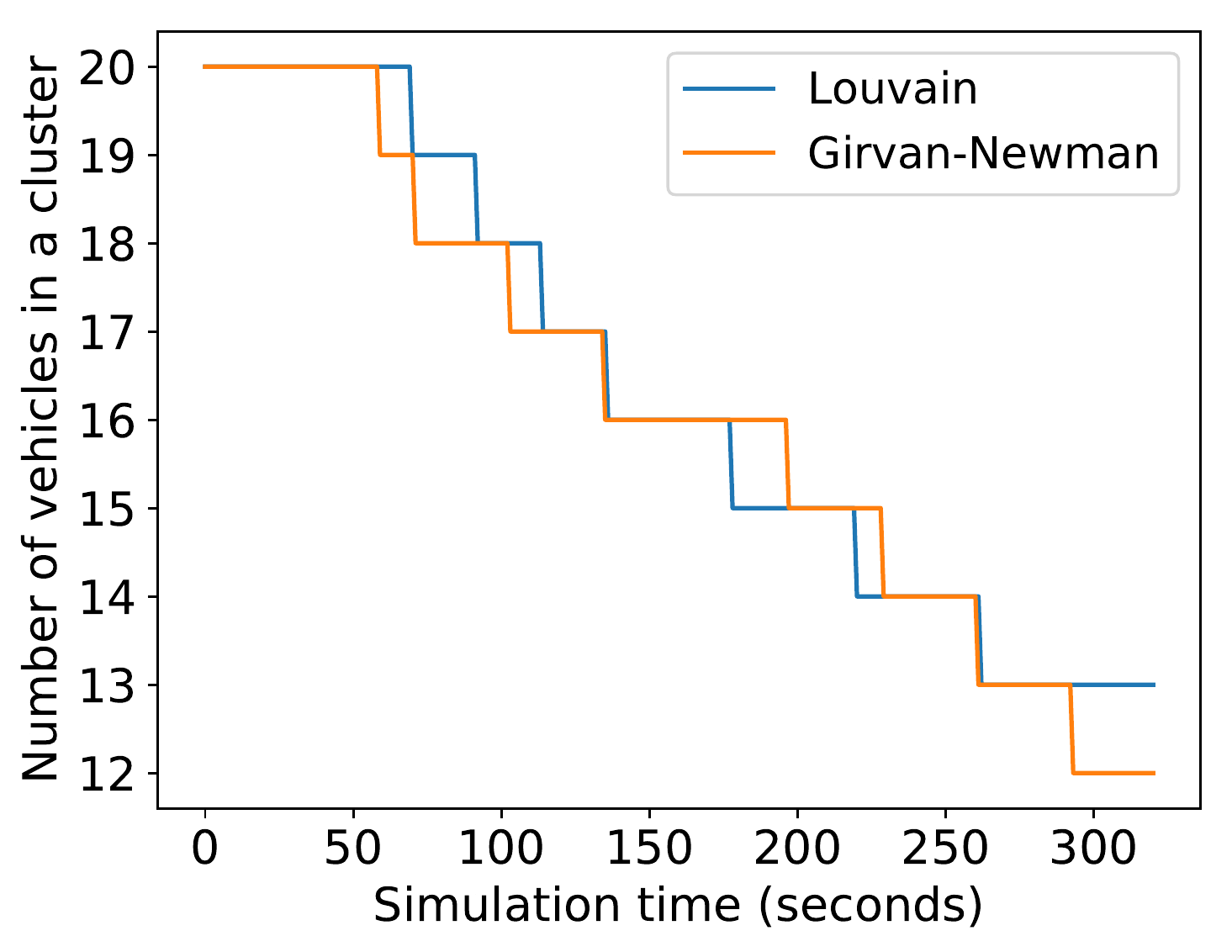}\vspace{-0.75em}
  \caption{\textbf{Cluster state through the simulation}}
  \label{fig:clusterperformance}
\end{subfigure}
\caption{Different performance metrics to evaluate the performance of the proposed heuristics, service time comparison for the proposed approach with edge placement and evaluation of the cluster state throughout the states of the simulation}
\label{fig:images}
\end{figure*}

As our heuristic solution does not select the shortest path but the most reliable path, it might select very long paths with multiple hops between the Type 1 and the CN TIs. We present both the aggregate hop count and the total number of scaled processing TIs (of Type 2, 3 and 4) corresponding to the number of Type 1 TI, presented in Fig.\ref{fig:hopcount}. For any number of TIs, the hop count of all the paths between Type 1 TI and CN ranges between 5 and 6. The total number of processing TIs for each placement is also plotted in the same figure and it ranges from 4 to 15.

To compare the number of scaled processing TI's, we run the optimization problem with the objective of minimizing the number of processing TIs, to analyze the least number of processing TIs required for meeting the application demands. We calculate the minimum number of Type 2, 3, and 4 TIs required for a successful service placement without any TI being rejected a placement on the vehicular cluster in Fig.\ref{fig:MinprocessingTI}. We compare this to the number of TIs scaled by our heuristic, plotted in Fig.\ref{fig:processingTI}. As shown, our heuristic scales are close to the minimum number of required TIs. It places 2-4 more TIs in comparison to the least number of required TIs. But the heuristic chooses more reliably connected nodes to place the TIs.

\subsection{Comparison of placement techniques in terms of service time}

We have evaluated our placement approach from a resource utilization point of view. We now look at a QoS-based metric to compare our approach to a mobile-edge computing-based solution. The service demands are still generated by moving vehicles, but the mobile-edge computing approach places all the TIs on static edge servers. For the real-time performance of the vehicle cluster, we use a fog computing simulator called Yet Another Fog Simulator (YAFS) \cite{8758823} for modeling the mobility and estimating the real-time performance of the selected cluster. YAFS is a python-based discrete event simulator that supports resource allocation policies in fog, edge, and cloud computing. The simulator has a distributed data flow application model that could be easily adapted to our use case. The simulation provides dynamic service selection, placement, and replacement of services that we have customized for our requirements. The support of mobility of nodes, which can also be treated as processing nodes makes the simulator a good fit for our case.

We consider the service time as the total time it takes for a service to execute, including both processing and link latency. We observe the minimum and maximum service time for the two services of different chain lengths for the mobile-edge placement versus our approach of placing multiple TIs on a moving vehicle cluster. We observe that the minimum service time is significantly less for the cloud placement, in comparison to our approach, in Fig.\ref{fig:Minservicetime}. Our approach places multiple TIs on different vehicles, thus the delay in the execution of one TI can result in a significant delay in service execution time. For the case of maximum service time, as can be seen in Fig.\ref{fig:Maxservicetime}, the mobile-edge placement is significantly high. This is because of the delay in sending all the collected data from moving vehicles to the edge server or RSU. The service time is also higher for the chain length of 4 for the mobile edge placement approach. In comparison, our approach approximately takes the same time for a chain length of 3 or 4 in the worst-case scenario as the minimum service time in Fig.\ref{fig:Minservicetime}.  Thus, even if an optimal placement is not achieved, on average our approach performs better in terms of service time, in comparison to the mobile-edge placement approach.


\subsection{Evaluation of the selected cluster over time}

We also analyze the performance of the selected cluster by evaluating the number of nodes in the cluster that stay together over a period of time.  For the two community detection-based node selection approaches, out of the 20 selected nodes, 12 to 14 nodes make it till the end of the simulation, as depicted in Fig.\ref{fig:clusterperformance}.

\begin{figure}[htb]
    \centering 
  \includegraphics[width=0.8\linewidth]{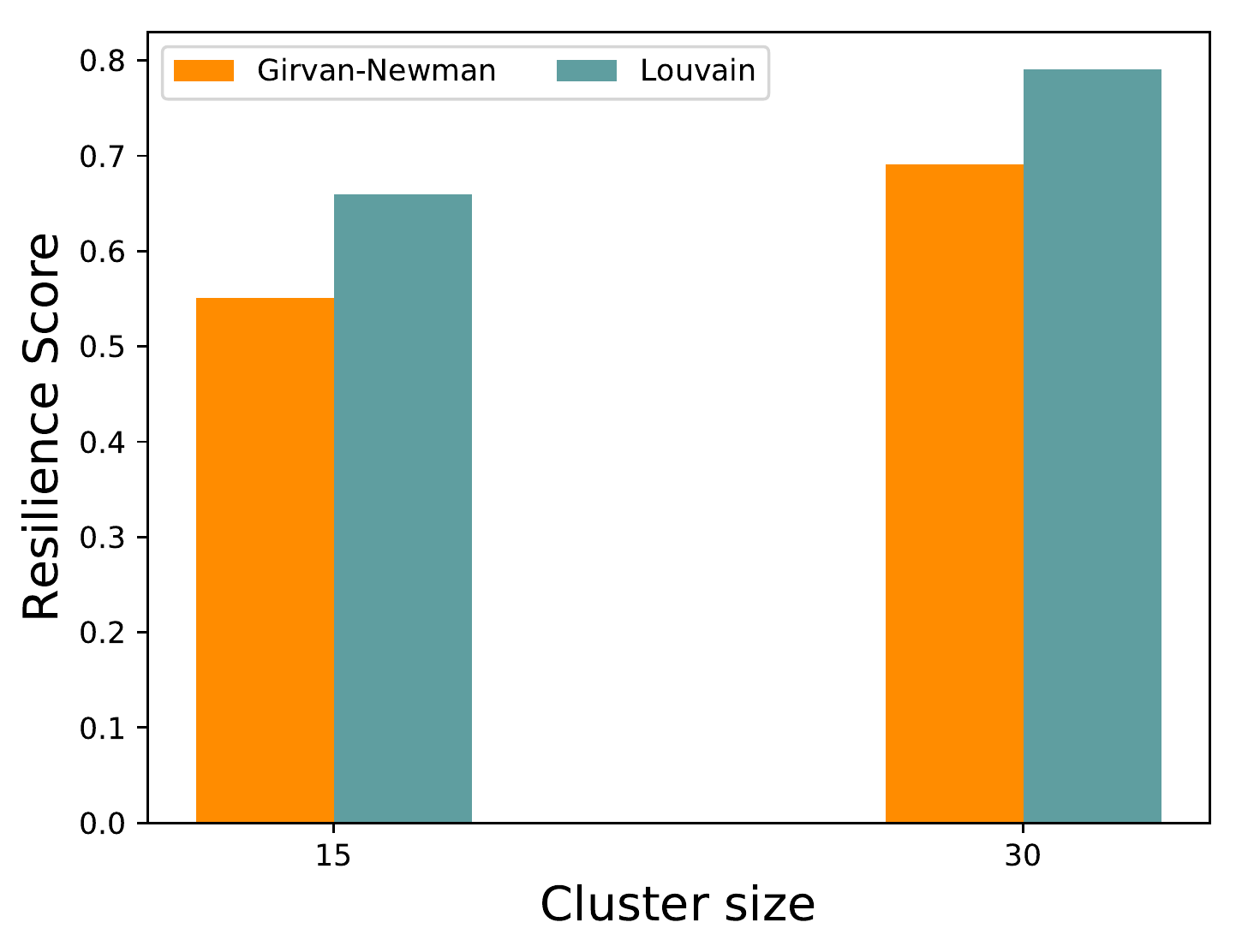}\vspace{-0.75em}
  \caption{Calculated betweenness centrality score for the two community-detection approaches}
  \label{fig:RS}
\end{figure}

We then evaluate the quality of the selected cluster in terms of the nodes that stay till the end of the simulation by using a resilience score. We use the betweenness centrality as a measure to check the importance of a node, in terms of connectivity in the graph. The \emph{betweenness centrality} calculates the shortest weighted path between every pair of nodes in a graph. Each node gets a betweenness centrality score (BCS) based on the number of shortest paths that pass through the node. The resilience score is calculated as the total BCS for all the nodes that made it till the end of the service time upon the total BCS of all the nodes in the selected cluster. The higher resilience score shows that nodes with higher BCS stayed with the cluster, thereby reducing the need for rerouting flows or re-configuring service chains due to the absence of a forwarding node or a path between two data-dependent TIs. We evaluate the communities detected for two community sizes, of 15 and 30, using the Girvan-Newman and the Louvain approaches. We observe a higher resilience score for the Louvain approach for communities of either size, as depicted in Fig.\ref{fig:RS}. 

We observe a higher resilience score for the bigger cluster, of 30 nodes, as we observe more number of nodes with higher BCS stay with the cluster till the end of the simulation time. However, this might not always be the case. A bigger cluster may not always end up being more resilient than a smaller cluster. The resilience score depends on the BCS, which is based on the importance or the role of a node in the graph in terms of the flow of communication. For example, the BCS of any vertex in a complete graph is zero since no vertex lies on the shortest path, as every node is connected to the other by a unique edge.

\section{Conclusion}

The paper aims to solve the problem of placing video collection and object-detection applications on moving vehicle clusters. The first application executes pre-processing tasks on the vehicle cluster whereas the second application executes a pre-trained local object-detection service, which is computation-intensive. We then model the problem as a multi-objective, constrained optimization problem. We introduce a node selection and service placement problem with the novelty of placing scalable and distributed services on mobile infrastructure. 

We also evaluate the performance of the service placement heuristic using other resource utilization measures like the number of scaled TIs and the average hop-count for placing the distributed services. We then consider a QoS-level parameter called service time to analyze how our approach performs compare to a mobile-edge placement approach. We also emulate the service placement using a Fog simulator. We analyze how the node selection approach performs in terms of the life of the selected cluster. We introduce a betweenness centrality-based resilience score to evaluate the performance of the chosen cluster, in terms of the quality of nodes that make it to the end of the execution time.

Our approach outperforms the integer linear program because it generates placement plans more quickly but with similar resource costs, and outperforms the baseline first-fit solution, because the mobility-aware strategy ensures that the cluster cohesion is higher, increasing the resilience of the system. We also compared our vehicular fog computing approach to edge computing, where the vehicles are not used for processing and just forward the data to the RSU and cloud for processing. Our placement technique results in better worst-case performance, with much lower maximum service time that is a measure of the time taken in service execution, including both processing and service latency.

In future work, we plan to extend our theoretical treatment of mobile service placement on vehicles in urban traffic by collecting data from a smart city testbed and analysing how well our algorithm performs in practice. We also intend to extend our theoretical treatment to consider service migration and not just the initial service placement problem. The revised model would consider migration costs and adding migration would make such services more robust when vehicles leave and their workload needs to be shared with other vehicles.




\bibliographystyle{IEEEtran}
\bibliography{vehicular}





\end{document}